\documentstyle[aps,preprint]{revtex}
\begin{document}
\draft

\title{
A Method to Study Relaxation of Metastable Phases:
Macroscopic Mean-Field Dynamics
}

\author{Jooyoung Lee,$^{1,*}$ M. A. Novotny,$^{1,\dag}$
and Per Arne Rikvold$^{1,2,\ddag}$}
\address{
$^1$ Supercomputer Computations Research Institute,\\
Florida State University, Tallahassee, FL 32306-4052\\
$^2$ Center for Materials Research and Technology, and
Department of Physics,\\
Florida State University, Tallahassee, FL 32306-3016
}

\date{\today}
\maketitle
\begin{abstract}
We propose two different macroscopic dynamics to describe the
decay of metastable phases in many-particle systems with local interactions.
These dynamics depend on the macroscopic order parameter $m$ through the
restricted free energy $F(m)$
and are designed to give the correct equilibrium distribution for $m$.
The connection between macroscopic dynamics
and the underlying microscopic dynamic are considered
in the context of a projection-operator formalism.
Application to the square-lattice nearest-neighbor
Ising ferromagnet gives
good agreement with droplet theory and Monte Carlo simulations
of the underlying microscopic dynamic.
This includes quantitative agreement for the exponential
dependence of the lifetime $\langle\tau\rangle$
on the inverse of the applied field $H$,
and the observation of distinct field regions in which
$\Lambda \equiv {\rm d} \ln \langle\tau\rangle / {\rm d}|H|^{1-d}$
depends differently
on $|H|$. In addition, at very low temperatures we observe
oscillatory behavior of $\Lambda$
with respect to $|H|$, due to the discreteness of the lattice and in
agreement with rigorous results.
Similarities and differences between this work and earlier works on finite
Ising models in the fixed-magnetization ensemble are discussed.
\end{abstract}


\section{Introduction}
\label{secint}

Metastable phases are observed in a wide variety of systems
that exhibit first-order phase transitions.
A few examples are supercooled fluids, permanent magnets, ferroelectrics,
certain alloys, the ``false vacuum''  associated with the electroweak
phase transition, and the supercooled quark/gluon plasma associated with the
QCD confinement transition.
In recent decades, much attention has been focused on the study of
metastable phases and the rate at which they decay
to thermodynamic equilibrium,
but a fully satisfactory description has remained elusive.
A recent review with numerous references to specific
realizations of metastable behavior in real and model
systems is found in Ref.~\cite{RG94}.

In certain systems with weak long-range interactions,
infinitely long-lived metastable phases can exist
in the thermodynamic limit \cite {PL}.
However, in systems with short-range interactions, there exist no
such stable non-equilibrium states, even in the thermodynamic limit.
Nevertheless, for large but finite systems,
the relaxation time for short-range models
can be extremely long compared with
any finite observation time \cite{Schulman,TM,RTMS,MAN}.
Here, we define the term ``metastability''
to include this phenomenon in short-range models.
The long relaxation time is mainly due to the large
free energy of the local fluctuations that must spontaneously
arise in order for the system to decay into a globally stable phase.
Due to the long relaxation time, it is difficult to tell
metastable phases from globally stable ones
by observing only short-time fluctuations.
The explorations of phase space, characteristic of the metastable phase,
are expected to be those included
in a constrained partition function that excludes
the microstates that dominate in equilibrium \cite{PL,FISHER}.
The application of such ideas to a field-theoretical droplet model
with Fokker-Planck dynamics has shown that close to coexistence,
the nucleation rate for droplets of the equilibrium phase
is proportional to the imaginary part of a complex-valued
constrained free energy obtained by analytic continuation
from the equilibrium phase into the metastable phase
\cite{Langer1,Langer2,Langer3}.

Recently, complex-valued constrained free energies
were numerically obtained for both the two-dimensional
nearest-neighbor Ising ferromagnet \cite{GRNa,GRNb}
and for models with weak long-range forces \cite{PBM_AIP,BPM94,FIIG}
by a constrained-transfer-matrix
method introduced by one of us \cite{PAR}.
Although no dynamical aspects were explicitly considered to
obtain the constrained free energies,
the average free-energy cost of a critical droplet was
obtained over a wide range of fields and temperatures,
in good agreement with the predictions of
field-theoretical droplet models
\cite{Langer1,Langer2,Langer3,BPM94,FIIG,GNW} and Monte Carlo (MC)
simulations \cite{RG94,TM,RTMS,MAN}.
These results indicate the relevance of
purely static properties, such as the free energy,
to the relaxation behavior of metastable phases.
Whereas any physical dynamic (consistent with
a real experimental situation) is bound to give the
correct {\it equilibrium} Boltzmann distribution
for an infinitely long observation time,
it is not yet clear how relevant the static properties
of a model are to the study of the
dynamical relaxation of a metastable phase towards equilibrium.

The observations
discussed in the preceding paragraph raise the interesting possibility
that the information stored in static quantities
may be sufficient to describe the salient features of
the relaxation behavior of a metastable phase, even in a short-range-force
system. A quantity which
contains all thermodynamically relevant equilibrium information
is the restricted bulk free energy,
\begin{equation}
F(m)=F_0(m)-\beta HNm \;,
\label{eqFdef}
\end{equation}
where $m$ is the macroscopic order parameter
conjugate to the external field $H$.
Here $\beta$$=$$1/T$ is the inverse temperature
with Boltzmann's constant $k_B$$=$$1$, and $\beta$ has been
absorbed in $F(m)$ and $F_0(m)$.
One can obtain $F(m)$ either exactly from exact enumerations
for small systems, or approximately, up to an additive constant,
by Monte Carlo simulation.
The importance of the detailed shape of
the zero-field free energy $F_0(m)$ for two-phase equilibria
and nucleation barriers has previously been discussed by
Binder and coworkers \cite{KB,FB,KASKI}.
Generalizations of Eq.~(\ref{eqFdef}) to consider several macroscopic
densities and their conjugate fields are straightforward.

Since the restricted bulk free energy is (by definition)
projected onto a space spanned by one
or a small number of macroscopic densities,
all detailed information about microscopic
spin configurations is lost.
In this paper we investigate the possibility that
macroscopic dynamical properties that are common
to several different microscopic dynamics may nevertheless be extracted from
the information retained in $F(m)$.
For this purpose, we construct two
different macroscopic, discrete-time dynamics, each defined
by a separate master equation for the order-parameter distribution
function.  Both master equations are
subject to the following two restrictions.\\
(1) The order parameter $m$ is allowed to change
only by a finite amount during each discrete time step.
(Locality in $m$.)\\
(2) The dynamics should reproduce the correct $F(m)$
in equilibrium. (Correct static properties of $F(m)$.)\\
Although the relevance to metastable decay of the
Hohenberg-Halperin scheme of dynamic universality classes
\cite{HOHA} is not completely clear, we demonstrate in this work
that the requirements (1) and (2) are sufficient to make our macroscopic
dynamics consistent with microscopic dynamics in the class of Model A:
systems with a nonconserved scalar order parameter and local dynamic.

For short-range-force models, the sequence of microscopic configurations
that constitutes a particular realization of a MC
simulation cannot be deduced from the corresponding sequence of
values of macroscopic variables. In contrast, for
models in which each site interacts equally with all other sites
while the total interaction energy remains independent of the system size
(equivalent-neighbor models), all configurations with identical values of the
order parameter are equivalent, so the dynamical properties can be
exactly obtained from the restricted free energy,
as has been shown by Griffiths, {\it et al.}~\cite{GWL}.
In the equivalent-neighbor limit, one of the macroscopic
dynamics that we propose in this work reduces
to the Metropolis \cite{METRO} version of the heat-bath dynamic studied
in Ref.~\cite{GWL}. (For discussions of the distinctions between
Metropolis and heat-bath or Glauber dynamics, see {\it e.g.}\
Refs.~\cite{RG94,Binder}.)
Since it is well known that both the equilibrium and the metastable
properties of equivalent-neighbor models are
exactly described by mean-field theory in the thermodynamic limit
\cite{PL,PBM_AIP,BPM94,FIIG}, these models are often
referred to as ``mean-field models'' \cite{GWL}.
Consistent with this usage, we
call the class of dynamics that we define here ``macroscopic mean-field
dynamics.''

Our proposed dynamics may be considered as approximations to the dynamic one
would obtain by projecting the microscopic dynamic onto
a master equation for the macroscopic order-parameter distribution,
using a projection-operator technique
\cite{NAKA58,ZWAN60,MORI65,GRAB77C,GRAB82,NORD75}.
This point of view is further explored in Appendix~\ref{app_pro_op}.
Since the decay of metastable phases is
a nonlinear, nonequilibrium problem, it is worth considering the extent to
which nonlinearities and correlations in the microscopic dynamic are included
in the proposed macroscopic dynamics.
We therefore point out that, although specifically {\it non}equilibrium
correlations are not included by virtue of the loss of spatial resolution
resulting from the projection of the dynamic onto the macroscopic
order parameter, {\it equilibrium}
correlations are included through their effect on the highly nonlinear
restricted free energy $F(m)$.
However, since in this paper we only consider a single macroscopic variable,
effects of nonlinear interactions between macroscopic variables \cite{GRAB82}
are not included.

The rest of this paper is organized as follows.
In Sec.~\ref{secmfd} we introduce our two specific macroscopic
mean-field dynamics, emphasizing their common physical motivation.
In Sec.~\ref{secrf} we show how to obtain $F(m)$ using MC simulations.
In Sec.~\ref{secdt} we summarize the relevant droplet-theory predictions
for the decay of metastable phases.
In Sec.~\ref{secnr} we apply our macroscopic dynamics
to the relaxation of the metastable phase
in the two-dimensional ferromagnetic nearest-neighbor Ising model
below its critical temperature.
In Sec.~\ref{secbind} we discuss the connections between our results for the
metastable lifetimes and the detailed shape of $F(m)$, with particular
reference to the earlier work by Binder and coworkers \cite{KB,FB,KASKI}.
Finally, in Sec.~\ref{secdis} we summarize our results and
discuss some implications of this study.

\section{Macroscopic Mean-Field Dynamics}
\label{secmfd}

To set the stage for our study, we first consider the
microscopic Metropolis dynamic for a mean-field Ising ferromagnet
in which each spin interacts with equal strength
with every other spin in the system.
As was already shown by Griffiths {\it et al.}\ \cite{GWL},
this exactly defines a macroscopic dynamic which depends on the
configurations of the system only through
the restricted free energy $F(m)$.
Next, for Ising models with finite interaction range,
we propose two macroscopic dynamics,
which can be constructed from $F(m)$ for the corresponding model
while satisfying the two conditions introduced in Sec.~\ref{secint}:
(1) locality in $m$ and (2) correct static properties of $F(m)$.
In proposing these dynamics we
make as few specific physical assumptions as possible
beyond the conditions (1) and (2).

The ferromagnetic Ising model with equivalent-neighbor interactions
is defined by the Hamiltonian
\begin{equation}
\label{efrs1}
{\cal H} = -(J/N) \sum_{i<j} s_i s_j
- H \sum_i s_i
\;,
\end{equation}
where $s_i$$=$$\pm 1$ are $N$ Ising spins,
$H$ is an external magnetic field, and
the sums $\sum_{i<j}$ and $ \sum_i$ run over all $N$ spins
with $1$$\le $$ i$$<$$j $$\le $$ N$ and $1$$\le $$ i $$\le $$ N$, respectively.
For convenience we set the interaction constant $J$ equal to 1.
The magnetization per spin,
\begin{equation}
\label{efrs1b}
m = N^{-1}\sum_i s_i \;,
\end{equation}
is the order parameter conjugate to $H$,
and the number of up spins is related to $m$ as
\begin{equation}
\label{efrs2}
n={N \over 2}(1+m)
\;.
\end{equation}
With these definitions, Eq.~(\ref{efrs1}) can be written as \cite{GWL}
\begin{equation}
\label{efrs3}
{\cal H}=E(n)=-{(2n-N)^2 \over 2N}-H(2n-N)+{1 \over 2}
\;.
\end{equation}

We consider a microscopic Metropolis dynamic
in which the spin at a randomly selected site $i$ is flipped
from $s_i \rightarrow -s_i$ with probability
\begin{equation}
\label{efrs4}
p(x \rightarrow x^\prime)  =
\exp \left[\min\left\{0,\beta\left(E(x)-E(x^\prime)\right)\right\}\right]
\;,
\end{equation}
where $E(x)$ and $E(x')$ are the energies of the microscopic
spin configurations $x$$=$$\{s_j\}$ and $x'$
before and after the flip, respectively.
The microscopic detailed-balance condition is satisfied \cite{Binder}.
For this model, the one-step transition probabilities
$W_1(n,n')$ from states with order parameter $n$ to
states with $n'$ are \cite{GWL}
\begin{mathletters}
\label{efrs5}
\begin{eqnarray}
\label{efrs5a}
W_1(n,n+1) & = &
\left(1-{n \over N}\right)
\exp \left[\min\left\{0,\beta \left(E(n)-E(n+1) \right)\right\}\right]\\
\label{efrs5b}
W_1(n,n-1) & = &
{n \over N}
\exp \left[\min\left\{0,\beta \left(E(n)-E(n-1) \right)\right\}\right]\\
\label{efrs5c}
W_1(n,n) & = & 1 - W_1(n,n+1) - W_1(n,n-1)
\;.
\end{eqnarray}
\end{mathletters}
Here the arguments of the matrix elements $W_1(n,n')$
are all between 0 and $N$, and matrix elements with arguments
outside this range are identically zero.
In this dynamic the probability for choosing an up (down) spin is
$n/N$ $(1-n/N)$.

Since the value of $n$ in the equivalent-neighbor model
uniquely specifies the energy of the spin configuration,
$\beta E(n)$ can be replaced by $F(n) + S(n)$,
where $F(n)$ is the restricted free energy
[$F(n) \equiv \beta E(n) - S(n)$ with $\beta$
absorbed in $F$],
and $S(n)=\ln \Omega(n)$ is the Boltzmann entropy for
the density of states $\Omega(n)= {N! / n! (N-n)!}$.
The probability that the system has $n$ up spins is
proportional to $\exp[-F(n)]$ \cite{LK}.
Eq.~(\ref{efrs5}) thus becomes
\begin{mathletters}
\label{efrs6}
\begin{eqnarray}
\label{efrs6a}
W_1(n,n+1) & = &
\left(1-{n \over N}\right)
\exp \left[\min\left\{0,\left(F(n)-F(n+1)+S(n)-S(n+1)\right)\right\}\right]\\
\label{efrs6b}
W_1(n,n-1) & = &
{n \over N}
\exp\left[\min\left\{0,\left(F(n)-F(n-1)+S(n)-S(n-1)\right)\right\}\right]\\
\label{efrs6c}
W_1(n,n) & = & 1 - W_1(n,n+1) - W_1(n,n-1)
\;.
\end{eqnarray}
\end{mathletters}
Since $\exp [S(n)-S(n-1)]=(N-n+1)/n$,
it is straightforward to show that
\begin{equation}
\label{efrs7}
{W_1(n,n^{\prime}) \over W_1(n^{\prime},n)}
= \exp [F(n)-F(n^{\prime})]
\;,
\end{equation}
which can be considered to be the macroscopic detailed-balance condition
between states with order parameter $n$ and $n^{\prime}$$=$$n$$\pm$$1$.
Consequently, the equilibrium probability distribution for
the order parameter $n$ resulting from this Metropolis dynamic
is correctly proportional to $\exp[-F(n)]$.

Once the transition probability matrix $W_1(n,n')$,
which for $N$ Ising spins is an $(N+1) \times (N+1)$
tridiagonal matrix, is constructed from $F(n)$,
particular realizations of this stochastic process
can be created, starting from an arbitrary initial state.
The average first-passage time
$\langle\tau\rangle$ to the globally stable phase, starting
from a metastable phase, can be easily calculated from $W_1(n,n')$
by the methodology of absorbing Markov chains  \cite{AMC}
(see Appendix~\ref{app_amc}).
Higher moments of the first-passage time $\tau$ can also be obtained
by the same method.
We again emphasize that for the equivalent-neighbor model, the time
evolution of the macroscopic order-parameter distribution given by
Eq.~(\ref{efrs6}) is an {\it exact\/} consequence of the underlying
microscopic dynamic \cite{GWL}.

Next, as a prototype ferromagnet with short-range interactions
we investigate the nearest-neighbor
square-lattice Ising model for which the Hamiltonian is
\begin{equation}
\label{nnIHam}
{\cal H} = - J \sum_{<i,j>} s_i s_j - H \sum_i s_i \;,
\end{equation}
where $s_i=\pm 1$ is the Ising spin at site $i$.
The interaction constant $J$ will be set to 1 as before,
periodic boundary conditions are used,
and $H$ is the applied field \cite{NOTE}.
The sums $\sum_{<i,j>}$ and $ \sum_i$ run over all
nearest-neighbor pairs and over all $N$$=$$L^2$ sites on a
square lattice.
In contrast to infinite-range models, for models with {\it finite}
interaction range, such as given by Eq.~(\ref{nnIHam}),
Eq.~(\ref{efrs6}) can not be exactly derived from the microscopic
Metropolis dynamic defined by Eq.~(\ref{efrs4}),
even if $F(n)$ is numerically calculated to give the correct
functional form appropriate to the particular model.
The reason for this is that the local environment
with which a particular spin interacts is no longer uniquely determined by
the macroscopic order parameter.
Nevertheless, we can define a macroscopic dynamic for a
particular short-range model
through Eq.~(\ref{efrs6}) with the appropriate form for $F(n)$ and consider
it as an approximation for the true order-parameter dynamic observed in a
microscopic MC simulation. This approximate macroscopic dynamic
is our ``mean-field dynamic No.~1'' (MFD1).
Although MFD1 is not exactly derived from any particular
microscopic dynamic, it satisfies the two conditions of
locality in the order parameter $m$
and correct static properties of $F(m)$, which we introduced in
Sec.~\ref{secint}.

We emphasize two important features of MFD1.
First, regardless of the functional form of $F(m)$
for the particular microscopic model,
this form is correctly reproduced by the dynamic.
Second, MFD1 is not the only macroscopic dynamic capable of
correctly yielding $F(m)$.
This is analogous to the fact that
there exist many different microscopic dynamics which can be successfully used
to study the equilibrium properties of a single model.
In fact, the macroscopic detailed-balance condition, Eq.~(\ref{efrs7}), is
sufficient to ensure that a dynamic yields $\exp[-F(m)]$ as
its equilibrium distribution. Perhaps the simplest Metropolis-type dynamic
which satisfies both Eq.~(\ref{efrs7}) and our requirement of locality in $m$,
is defined by the transition-matrix elements
\begin{mathletters}
\label{efrs9}
\begin{eqnarray}
\label{efrs9a}
W_2(n,n+1) & = &
 {1 \over 2} \exp [\min\{0,F(n)-F(n+1)\}] \\
\label{efrs9b}
W_2(n,n-1) & = &
 {1 \over 2} \exp [\min\{0,F(n)-F(n-1)\}] \\
\label{efrs9c}
W_2(n,n) & = & 1 - W_2(n,n+1) - W_2(n,n-1)
\;,
\end{eqnarray}
\end{mathletters}
where the range of the arguments is the same as for $W_1$,
and $W_2$$=$$0$ for out-of-range arguments.
These transition probabilities define our ``mean-field dynamic No.~2'' (MFD2).

In Sec.~\ref{secnr} we compare
the macroscopic order-parameter dynamic observed in MC
simulations of the two-dimensional Ising ferromagnet with the
microscopic Metropolis dynamic defined by
Eq.~(\ref{efrs4}) to our approximate macroscopic dynamics, MFD1 and MFD2.
This allows us to investigate the relevance of the
equilibrium properties of a system to the dynamical relaxation
behavior of its metastable phases.

\section{Restricted free energy by Monte Carlo simulations}
\label{secrf}

In this section we describe how to obtain, from microscopic, equilibrium
MC simulations, the bulk restricted free energy $F(m)$, which is defined
through the restricted partition function
\begin{equation}
\label{efrs15}
\exp [-F(m)]= \sum_x \delta(m(x)-m) \exp [-\beta E(x)]
 \;,
\end{equation}
where the sum is over all possible {\it microscopic}
spin configurations $x$,
and $\delta$ is the Kronecker delta function.
It is straightforward to show that the
probability distribution for $m$ is proportional to $\exp [-F(m)]$ \cite{LK}.
For Ising models with short-range interactions below the critical temperature
$T_{\rm c}$, $F_0(m)$ ({\it i.e.\/} $F(m)$ for $H$=0, as defined in
Eq.~(\ref{eqFdef})) has two symmetrical minima,
and the bulk free-energy barrier separating these minima
diverges as the linear system size $L$ for $L$$\gg$$1$ in
two dimensions (as $L^{d-1}$ in $d$ dimensions) \cite{LK}.

Recently, significant progress has been achieved
in the search for more efficient MC sampling algorithms
for systems in which different subsets of phase space are
separated by large free-energy barriers
\cite{VC,TV,KW,Huller,BN,BCB,BHN,Lee}.
Here we use a variation of the multicanonical method
\cite{BN,BCB,BHN}, employing the notation of Ref.~\cite{Lee}.

The partition function for $H$=$0$ and inverse temperature $\beta$ is
\begin{equation}
\label{efrs16a}
Z(\beta,H$=$0) = \sum_{E,M}
\exp [S(E,M)-\beta E] = \sum_{M} \exp [-F_0(M)]
\;,
\end{equation}
where
\begin{equation}
\label{efrs16b}
\exp [-F_0(M)] \equiv \sum_{E} \exp [S(E,M)-\beta E]
\;,
\end{equation}
$\exp [S(E,M)]$$=$$\Omega(E,M)$ is the density of states,
and $E$ and $M$$=$$mL^2$ are the bulk internal energy and magnetization,
respectively.

The detailed-balance condition for the MC simulation
can be written as
\begin{equation}
\label{efrs17}
{W(x \rightarrow x^{\prime}) \over W(x^{\prime} \rightarrow x)}
= \exp\left[-\beta\left\{E( x^{\prime})-E(x)\right\}-
\left\{\widetilde{J}\left(M(x^{\prime})\right)
-\widetilde{J}\left(M(x)\right)\right\}\right]
\;,
\end{equation}
where $\widetilde{J}$ can be any arbitrary function of $M$.
For an ergodic MC algorithm, the resulting distribution (histogram)
of the sampling has been shown \cite{Lee} to be
\begin{equation}
\label{efrs18}
\widetilde{H}(E,M) \propto \exp[S(E,M)-\beta E-\widetilde{J}(M)]
\;.
\end{equation}
{From} Eqs.~(\ref{efrs16b}) and (\ref{efrs18}) we get
\begin{equation}
\label{efrs19}
\widetilde{H}(M) = \sum_{E} \widetilde{H}(E,M)
     \propto\exp [F_0(M)-\widetilde{J}(M)]
\;.
\end{equation}
Therefore,
\begin{equation}
\label{efrs20}
F_0(M) =
  \widetilde{J}(M) + \ln \widetilde{H}(M)
\end{equation}
up to an additive constant.

As in Ref.~\cite{Lee}, the quantity $F_0(M)$ can be obtained
using Eq.~(\ref{efrs20}) in an iterative fashion.
The old estimates of $F_0(M)$ and $\widetilde{J}(M)$
can be used in a MC procedure to obtain a new histogram
and consequently a new and better estimate for $F_0(M)$
from Eq.~(\ref{efrs20}).
Although this approach is more efficient than
conventional sampling methods \cite {BHN},
we find that it does not suffice
to obtain $F_0(M)$ at very low temperatures.
This is due to the large size of the exponent
in Eq.~(\ref{efrs17}) for large $\beta$.

As in Ref.~\cite{Lee}, we instead sample the Boltzmann entropy
$S(E,M)$ directly by imposing the detailed-balance condition
\begin{equation}
\label{efrs21}
{W(x \rightarrow x^{\prime}) \over W(x^{\prime} \rightarrow x)}
= \exp\left[\widetilde{J}\{E(x),M(x)\}-\widetilde{J}\{E(x'),M(x')\}\right]
\;,
\end{equation}
which yields the histogram
\begin{equation}
\label{efrs22}
\widetilde{H}(E,M) \propto \exp[S(E,M)-\widetilde{J}(E,M)]
\;.
\end{equation}
Therefore, $S(E,M)$ can be obtained, up to an additive constant,
by iteratively evaluating
\begin{equation}
\label{efrs23}
S(E,M)=
  \widetilde{J}(E,M) + \ln \widetilde{H}(E,M)
\end{equation}
where $\widetilde{J}(E,M)$ is an input and $\widetilde{H}(E,M)$ is the
resulting histogram of $\widetilde{J}(E,M)$ from Eq.~(\ref{efrs21}).
Once $S(E,M)$ has been obtained to the desired accuracy,
$F_0(M)$ can be obtained
from  Eq.~(\ref{efrs16b}) for any $\beta$.
We were able to obtain $F_0(M)$ for any $\beta $$>$$ \beta_{\rm c}$
for the nearest-neighbor Ising ferromagnet on $L$$\times$$L$
square lattices with $L$$\le$$24$.
The $H$-dependent $F(m)$ are trivially constructed from $F_0(m)$ by
Eq.~(\ref{eqFdef}).

In Fig.~\ref{figfm24} we show $F_0(m)$ for
$T$$=$$0.8T_{\rm c} \approx 1.8153$ (Fig.~\ref{figfm24}(a))
and for $T$$=$$0.2 \approx 0.0881 T_{\rm c}$ (Fig.~\ref{figfm24}(b)).
The sawtooth-like behavior of $F_0(m)$ for $T$$=$$0.2$ is due to the
discreteness of the lattice.
For $T$$=$$0.8T_{\rm c}$ we used Eq.~(\ref{efrs17})
to obtain $F_0(m)$ for lattices up to $L$$=$$64$.
The vertical arrows marked $m_{\rm c}$
indicate the exactly known magnetization
for which the most likely  configuration with the given magnetization
changes from a slab (for $|m| < m_{\rm c}$)
to a droplet (for $|m| > m_{\rm c}$)
in an infinite system \cite{LZ}. In Fig.~\ref{figfm24}(a) we have also
marked the value of $m_{\rm c}$ corresponding to $L$=24, taken from
Fig.~3 of Ref.~\cite{LZ}. For further discussion of the droplet-to-slab
transition and its significance, see Sec.~\ref{secbind}.

\section{Droplet Theory and its Predictions}
\label{secdt}

In this section we review some predictions of droplet theory
for the metastable lifetime $\langle\tau\rangle$ in $d$-dimensional
ferromagnetic systems with short-range interactions and local dynamics
\cite{NOTE}. We leave details of the theory to Refs.~\cite{RG94,RTMS}
and the references cited there.
Starting with a large magnetization opposite to the applied field, we
consider its relaxation and define
$\langle\tau\rangle$ as the average time it takes for $m(t)$
to reach a particular cutoff value, $m_{\rm cut}$.

In addition to the lattice constant, which we take as unity,
five length scales are important in the droplet theory for relaxation
of metastable phases. These are the linear system size $L$,
the radius of the critical droplet $R_{\rm c}$,
the average distance between supercritical droplets $R_0$, and the
single-phase correlation lengths in the stable and metastable
phases, $\xi_{\rm s}$ and $\xi_{\rm ms}$, respectively.
Since the temperatures of interest in the present study are well below
$T_{\rm c}$, $\xi_{\rm s}$ is practically field independent and of the order
of unity, and since the field strengths considered are moderate,
the same is the case for $\xi_{\rm ms}$ \cite{GRNb}. Thus we are left
to consider the interplay between three lengths: $L$, $R_0$, and
$R_{\rm c}$, all of which are larger than unity in the temperature- and
field regimes of interest.

By comparing the bulk free energy gained by
creating a droplet of the equilibrium phase in the metastable
background with the free-energy cost of creating the droplet surface,
one can show that  the critical radius, beyond which the droplet
is more likely to grow than to shrink, is
\begin{equation}
\label{eqRc}
R_{\rm c} = \frac{(d-1) \sigma_0(T)}{ \Delta m |H|}
\approx \frac{(d-1) \sigma_0(T)}{2 m_{\rm eq}(T) |H|}
\,
\end{equation}
where
$\Delta m$ is the magnetization difference between the
metastable and the stable phase, and $m_{\rm eq}(T)$ is the spontaneous
equilibrium magnetization.
Also, $\sigma_0(T)$ is the equilibrium surface tension along a primitive
lattice vector, and is assumed to be equal to the surface
tension in the metastable phase.
For the two-dimensional Ising model, both
$\sigma_0(T)$ \cite{ZA} and $m_{\rm eq}(T)$ \cite{CNY} are exactly known.
The relaxation proceeds in different ways,
depending on the relative sizes of $L$, $R_0$, and $R_{\rm c}$.

If $|H|$ is sufficiently small that $R_{\rm c}$$>$$L$, then the
saddle-point configuration is a slab spanning the system in $d \! - \! 1$
dimensions, and $L$ is the most important length scale for the relaxation.
The metastable lifetime is then determined by the
surface free energy of such a slab.  Consequently, for periodic
boundary conditions it
increases with $L$ as \cite{BIND81,BIND82,BERG93}
\begin{equation}
\label{eq:celife}
\langle \tau(T,H,L) \rangle
\sim \exp \left( 2 \beta \sigma_0(T) L^{d-1} \right) \ .
\end{equation}
This region of ultraweak fields is called the
``coexistence'' (CE) region \cite{RTMS}, since the
dynamic is similar to that at $H$$=$$0$,
where two competing bulk phases coexist.

As $|H|$ is increased, $R_{\rm c}$ becomes smaller than $L$.
The crossover to the field regimes where the relaxation is dominated by
critical droplets smaller than $L$ has been called the
``thermodynamic spinodal'' (THSP) \cite{TM}.
The crossover field $H_{\rm THSP}$
can be estimated by requiring that the critical
droplet should occupy a volume fraction
\begin{equation}
\label{eq:phidef}
\phi = \left( 1 \! - \! \frac{m}{m_{\rm eq}(T)} \right)
\end{equation}
corresponding to the cutoff magnetization $m_{\rm cut}$.
This yields \cite{RG94,RTMS}
\begin{equation}
\label{eq:hthsp}
H_{\rm THSP}(\phi) = \frac{1}{L}
\left[ \frac{(d \! - \! 1) \Xi(T) }{ 2 m_{\rm eq}(T) \phi} \right]^{1/d}
\;,
\end{equation}
where $\Xi(T)$ can be calculated from $m_{\rm eq}$
and the anisotropic equilibrium surface tension by
the equilibrium Wulff construction \cite{ZA,RW}
to obtain the critical droplet shape \cite{GRNa,GRNb,NOTE3}.

{For} fields somewhat stronger than $H_{\rm THSP}$, so that
$L$$\gg$$R_{\rm c}$$\gg$$1$,
the nucleation rate per unit volume for critical droplets becomes
\cite{Langer1,Langer2,Langer3,GNW}
\begin{equation}
\label{efrs24}
\Gamma (T,H) \propto  |H|^{b+c}
\exp \left[{-\beta\Xi(T)\over |H|^{d-1}}\left\{1+O(H^2)\right\}\right]
\;,
\end{equation}
The exponent $b$ is a universal
exponent related to excitations on the droplet surface,
and  the nonuniversal exponent $c$ gives the $H$ dependence of
a ``kinetic prefactor'' \cite{Langer1,Langer2,Langer3}
which contains all dependence on the details of the dynamics. For
$d$=2 and~3 it is expected that $b$$=$$1$ and~$-7/3$, respectively
\cite{GNW}. This has been confirmed by several methods, most recently by
constrained-transfer-matrix calculations \cite{GRNa,GRNb}.
For dynamics that can be described by a Fokker-Planck equation
it is expected that $c$$=$$2$ \cite{Langer2,Langer3,GNW}.
In a recent MC study for $d$=2 it was confirmed that $b$+$c$$\approx$3 for
the Metropolis dynamic with updates at randomly chosen sites \cite{RTMS}.

If $R_0$$\gg$$L$$\gg$$R_{\rm c}$,
a single critical droplet is sufficient for macroscopic decay
to occur before additional droplets nucleate.
This region is called the ``single-droplet'' (SD) region \cite{RTMS}.
Assuming that the exact value of $m_{\rm cut}$ is not an important
factor (see the discussion in Ref.~\cite{RTMS}),
the average relaxation time
$\langle\tau\rangle$ to $m_{\rm cut}$ can be written as
\begin{equation}
\label{efrs25}
\langle\tau(T,H,L)\rangle \approx
\left [ L^{d} \Gamma (T,H) \right ]^{-1}
\propto  L^{-d} |H|^{-(b+c)}
\exp \left[{\beta\Xi(T)\over |H|^{d-1}}\left\{1+O(H^2)\right\}\right]
\;.
\end{equation}
This leads to
\begin{equation}
\label{efrs26}
\Lambda (T,H) \equiv
{{{\rm d} \ln \langle\tau(T,H,L)\rangle} \over {{\rm d} |H|^{1-d}}}
= \beta\Xi(T) + {b \! + \! c \over d \! - \! 1 } |H|^{d-1}
\;,
\end{equation}
where we have neglected higher-order correction terms.

If $L$$\gg$$R_0$$\gg$$R_{\rm c}$,
many critical droplets nucleate before the decay of the
order parameter can proceed to a macroscopic extent.
This region is called the ``multi-droplet'' (MD) region \cite{RTMS}.
In this region, we expect $\langle\tau\rangle$ to be independent of $L$.
With the assumption that the radial growth velocity
of the supercritical droplets is proportional to the applied field $H$
\cite{LIFS62,CHAN77,ALLE79},
$\langle\tau\rangle$ is predicted to be
\cite{RTMS,KOLM37,JOHN39,AVRAMI,SEKIMOTO1,SEKIMOTO2}
\begin{equation}
\label{efrs27}
\langle\tau(T,H)\rangle
\sim |H|^{-{b+c+d \over d+1}}
\exp\left[{\beta\Xi(T) \over (d \! + \! 1)
|H|^{d-1} }\left\{1+O(H^2)\right\}\right]
\;,
\end{equation}
which leads to
\begin{equation}
\label{efrs28}
\Lambda(T,H)
={ \beta\Xi(T)\over d \! + \! 1}
+ {b \! + \! c \! + \! d \over d^2 \! - \! 1} |H|^{d-1}
\;,
\end{equation}
where higher-order corrections again have been neglected.
One also has the mean droplet distance \cite{RTMS,SEKIMOTO1,SEKIMOTO2}
\begin{equation}
\label{efrs29}
R_0(T,H) \propto
|H|^{-{b+c-1 \over d+1}}
\exp\left[{\beta\Xi(T) \over (d \! + \! 1) |H|^{d-1}}
 \left\{1+O(H^2)\right\}\right]
\;.
\end{equation}
The field $H_{\rm DSP}$,
which separates the SD and MD regions was
called the ``dynamic spinodal field'' in Refs.~\cite{TM,RTMS}.
It can be estimated by setting
$R_0(T,H)$$\propto$$L$ with a proportionality constant of order unity,
which gives the scaling relation,
\begin{equation}
\label{efrs30}
H_{\rm DSP} =
\left( \frac{\beta \Xi(T)}{(d \! + \! 1) \ln L}\right)^{1 \over d-1}
\left[ 1 + O\left( \frac{\ln ( \ln L) }{\ln L } \right)
         + O\left( \frac{1}{\ln L } \right) \right]
\; ,
\end{equation}
expected to be asymptotically valid for nonzero $T$.

These droplet-theoretical results can be summarized as follows.
{For} a  given system size $L$, if $|H|$ is
small enough so that $R_{\rm c}$$>$$L$,
then $L$ is the most important length scale for the relaxation, and the
system is in the CE region.
As $|H|$ is increased beyond $H_{\rm THSP}$,
$R_{\rm c}$ becomes smaller than $L$.
If $R_0$$\gg$$L$$\gg$$R_{\rm c}$, the system is in the SD region,
and the relaxation is characterized by Eq.~(\ref{efrs26}),
from which one can estimate $b$$+$$c$ and $\Xi(T)$.
As $|H|$ is increased still further, so that
$R_{\rm c}$$\ll$$R_0$$\ll$$L$, the system is in the MD region.
The relaxation is then characterized by Eq.~(\ref{efrs28}),
from which one again can estimate $b$$+$$c$ and $\Xi(T)$.
The crossover field that separates the SD and MD regions,
$H_{\rm DSP}$, also separates the ``stochastic region'' and
the ``deterministic region'' \cite{TM}.
In the ``deterministic'' region, the average
metastable lifetime $\langle\tau\rangle$
is quite short, and the standard deviation of $\tau$ is
much smaller than $\langle\tau\rangle$.
In the ``stochastic'' region, the decay approximately follows a
Poisson process, so that the standard deviation of $\tau$
is comparable to $\langle\tau\rangle$.
As $|H|$ becomes very large, the droplet
picture becomes inappropriate, and
$R_0$ and $R_{\rm c}$ become comparable to the lattice spacing.
The lifetime is then on the order of one MC step per site (MCSS),
and this region is called the ``strong-field'' (SF) region.
The crossover between the MD and SF regions is marked by the
``mean-field spinodal field'' $H_{\rm MFSP}$ \cite{TM}, which can be
estimated as the field at which $2 R_{\rm c}$=1 \cite{GRNb}.
In Fig.~\ref{fig1O} we show a schematic diagram for the relaxation behavior,
illustrating the four sub-regions of characteristic
relaxation behavior predicted by droplet theory.

{For} low temperatures
the discreteness of the lattice becomes important.
It has been shown \cite{NS,SCH,MOS,Sco1,KO,Sco2}
that for sufficiently low temperatures
the lifetime of the metastable phase for the nearest-neighbor
square-lattice Ising ferromagnet with Hamiltonian given by
Eq.~(\ref{nnIHam}) is given by
\begin{equation}
\label{efrs40}
\ln\langle L^2 \tau \rangle
=8 \beta l_{\rm c}-2 \beta |H| (l_{\rm c}^2-l_{\rm c}+1)
\;,
\end{equation}
where the size of the critical droplet is
$l_{\rm c}= \lceil 2/|H| \rceil$,
and the notation $\lceil x \rceil$
denotes the smallest integer greater than $x$.
This result is restricted to $2/|H|$ not being
an integer and to $|H|<4$.  For this result to be valid the
temperature must be at least low enough and the lattice size large
enough to insure that the system is in the SD region.
Differentiating Eq.~(\ref{efrs40}) with respect to $|H|^{-1}$ gives
\begin{equation}
\label{efrs41}
T \Lambda  = 2H^2(l_{\rm c}^2-l_{\rm c}+1)
\;.
\end{equation}

\section{Numerical Results}
\label{secnr}

In this section we present extensive results of the mean-field dynamics
introduced in Sec.~\ref{secmfd} for
the relaxation behavior of the metastable phase
in the nearest-neighbor Ising ferromagnet on a square lattice
with the Hamiltonian given in Eq.~(\ref{nnIHam}).  Periodic
boundary conditions are used throughout.

First, we consider $T$$=$$0.8T_{\rm c}$ $(\beta$$=$$0.55086$$\dots$$)$
for direct comparison with recent MC results \cite{RTMS}.
We obtained $F_0(m)$ for the
entire range of $-1 \le m \le +1$ for system sizes up to $64$$\times$$64$,
using the method outlined in Sec.~\ref{secdt} and utilizing
the Ising spin-reversal symmetry $F_0(m)=F_0(-m)$.
For $H \ne 0$, $F(m)$ was obtained from Eq.~(\ref{eqFdef}).
Once $F(m)$ was obtained, the Markov transition probability matrices
${\bf W}_1$ and ${\bf W}_2$ for the two dynamics MFD1 and MFD2
were constructed through the procedure outlined in Sec.~\ref{secmfd}.
We show the results from MFD1 first and discuss MFD2 later.
As in Ref.~\cite{RTMS} the initial state is chosen as
$m$$=$$+1$ with $H$$<$$0$,
and an absorbing barrier is put at $m_{\rm cut}$$=$$0$.
In Ref.~\cite{RTMS} different values of $m_{\rm cut}$ were also used.
As long as $m_{\rm cut}$ is sufficiently far away from the metastable
value of $m$ that the largest droplet must already be supercritical,
the precise value of $m_{\rm cut}$ is not important
for weak fields \cite{RTMS}.

Using the absorbing Markov chain method
discussed in Appendix~\ref{app_amc},
the average first-passage time $\langle\tau\rangle$
to $m$$=$$m_{\rm cut}$ and the standard deviation
$\sigma_{\tau}$$\equiv$$\sqrt{\langle\tau^2\rangle-\langle\tau\rangle^2}$
were obtained by matrix inversion using the subroutine
{\tt tridag} from Ref.~\cite{NR}.
Our values of $\tau$ were divided by $L^2$
to give all times in units of MCSS.

In Fig.~\ref{fig2O} we show the  relaxation time,
defined as the first-passage time to $m_{\rm cut}$$=$$0$,
obtained from MFD1 for $L$$=$$32$ and 64.
For comparison, MC results for the standard Metropolis algorithm
with spin updates at randomly chosen sites for $L$$=$$32$ and $64$
are also plotted.
We find that MFD1 and the microscopic MC simulations give
qualitatively similar results for the relaxation time.

In Fig.~\ref{fig3O} we show the slope, $\Lambda$, of the data in
Fig.~\ref{fig2O} with respect to $|H|^{-1}$,
whose asymptotic values in the SD and MD regions are given in
Eqs.~(\ref{efrs26}) and (\ref{efrs28}), respectively.
We clearly see four distinct relaxation regions for $L$$=$$64$.
The SF region, $|H| \! > \! H_{\rm MFSP}$
($H_{\rm MFSP}(0.8T_{\rm c}) \! \approx \! 0.75$ \cite{GRNb}), contains the
sharp peak for very small $|H|^{-1}$ in Fig.~{\ref{fig3O}}(b).
(Fig.~{\ref{fig3O}}(a) is analogous with the schematic
Fig.~{\ref{fig1O}, in which the
various field regions and crossover fields are indicated.)
The region $0.2$$\lesssim$$|H|$$\lesssim$$0.75$
corresponds to the MD region.
As $|H|$ is lowered, the crossover from the MD region to
the SD region is signalled by a sudden rise in $\Lambda$.
In Fig.~{\ref{fig3O}}(a)
we define $H_{\rm max}$ and $H_{\rm min}$ as the fields
at which $\Lambda$ has a local maximum and minimum, respectively.
As $|H|$ is lowered further,
$\Lambda$ plunges towards zero, signalling the CE region. From
our numerical data for MFD1 we find this crossover field
$H_{\rm THSP}$ at about $|H|^{-1}\approx 40$ for $L$$=$$64$
and $|H|^{-1}\approx 20$ for $L$$=$$32$. (See Fig.~{\ref{fig3O}}(b).)
This is consistent with the relation $H_{\rm THSP} \propto L^{-1}$
given in Sec.~\ref{secrf}.
For further discussion of the thermodynamic spinodal and its
relation to the droplet-to-slab transition, see Sec.~\ref{secbind}.

Using the numerically exact value \cite{ZA,RW} of $\Xi$, we find
our results consistent with $b$$+$$c$$\approx$$2$ for MFD1,
based on the data in the SD region.
The expected value for dynamics that can be described by a Fokker-Planck
equation is $b$$+$$c$$=$$3$.
The estimate $b$$+$$c$$\approx$$2$ for MFD1 at $0.8$$T_{\rm c}$
should be taken with extreme caution, since
the asymptotic region for $L$$=$$64$ appears to be quite small.
For a more reliable estimate,
we would need results from larger systems at this temperature.
Since we have difficulties in obtaining $F(m)$ for larger
system sizes, we instead tested Eq.~(\ref{efrs26})
at lower temperatures,
assuming that $b$$+$$c$ does not depend on the temperature.
Although the accessible system sizes are smaller
for lower temperatures, due to the difficulties in estimating
$S(E,M)$ for all values of $E$ and $M$,
this approach turns out to be a more reliable
way with our dynamics to estimate $b$$+$$c$.  These studies at lower
$T$ are also consistent with $b$$+$$c$$\approx$$2$, as discussed below.

Although the results from the mean-field dynamic are
in good overall agreement with those from the MC simulations,
the quantitative estimates of the intercept and the
slope in Fig.~{\ref{fig3O}}(a) are
less satisfactory for the MD region.
Since the asymptotic region, in which Eq.~(\ref{efrs28}) is valid for
MC simulations at 0.8$T_{\rm c}$,
has been shown to be relatively narrow, even for $L$$=$$720$
\cite{RTMS}, results for much larger system sizes than $L$$=$$64$
would be necessary to provide a satisfactory test of Eq.~(\ref{efrs28})
for the mean-field dynamics at this temperature.
At lower temperatures we are nevertheless able to estimate below that
$b$$+$$c$$\approx$$2$ for MFD1 in the SD region, and
this gives the slopes drawn in Fig.~{\ref{fig3O}}(a) for both regions.
In contrast, it was demonstrated in Ref.~\cite{RTMS} for systems with
$L \! \le \! 720$
that Metropolis MC with spin updates at randomly
selected sites gives $b$$+$$c$$\approx$3,
in agreement with theoretical expectations
\cite{Langer2,Langer3,GNW}. The apparent
consistency with $b$$+$$c$$\approx$$2$
of the Metropolis MC results for $L$$=$$64$, which are also shown
in Fig.~{\ref{fig3O}}(a), is therefore clearly due to finite-size effects.

In Fig.~\ref{fig5O} we show the field dependence of the slope $\Lambda$
obtained from MFD1 and MFD2 at $T$$=$$0.8T_{\rm c}$.
Both are in qualitative agreement with Fig.~\ref{fig1O}.
The SF region, where $\Lambda$ decreases to zero for large $|H|$,
is not shown.
Figure~\ref{fig5O} provides some insight about
the relevance of $F(m)$ and the possible artificial
results from the mean-field dynamics.
The fact that Fig.~\ref{fig5O} is in qualitative
agreement with Fig.~\ref{fig1O}
indicates the importance of $F(m)$ for the dynamics.
However, as one might expect from the sensitive dependence of the
kinetic-prefactor exponent $c$ on the details of the dynamic,
which was demonstrated in Ref.~\cite{RTMS},
values of $b$+$c$ obtained from mean-field dynamics do not correctly
reflect that of the underlying microscopic dynamic.
In the SD region, $\Lambda$ approaches the exact value $\beta\Xi$
with a slope that depends on the details of the particular dynamic.
For MFD1 and MFD2, $b$$+$$c$ is about 2.0 and 3.7, respectively.
In the MD region for both MFD1 and MFD2, we find only
qualitative agreement with droplet-theory predictions.
As will be explained in Sec.~\ref{secbind}, it is expected that
only in the SD region can quantities related to the metastable
phase be reliably extracted from $F(m)$.
In the MD region $\Lambda$ depends
strongly on the details of the particular dynamic, and the
dynamics associated with the
droplet growth is not taken into account correctly in $F(m)$.
Note that the MD region is fairly easily accessible by standard MC simulations,
since $\tau$ there is rather small.

In Fig.~\ref{fig6O} we show the field dependence of the relative standard
deviation of the lifetime,
\begin{equation}
r=\sigma_{\tau}/\langle\tau\rangle
\;,
\end{equation}
which provides additional information about
how the metastable phase decays.
If the decay of the metastability
involves a single Possion process of forming one critical droplet,
as is the case in the SD region, we expect $r$$\approx$$1$.
In the MD region, on the other hand,
one needs to consider many independent Poisson processes.
By partitioning the system into $(L/R_0)^2$$\gg$$1$
cells of volume proportional to $R_0^2$, one gets $r\propto R_0/L$,
where $R_0$ is given by Eq.~(\ref{efrs29}) \cite{RTMS,RICH95}.
A more rigorous argument, based on the two-point correlation
function \cite{SEKIMOTO2}, can be found in the appendix of Ref.~\cite{RICH95}.
An estimate for the crossover field $H_{\rm DSP}$
between the SD and MD regions can be chosen as the field
$H_{1/2}$, for which $r$$=$$1/2$ \cite{TM,RTMS}.
For further discussion of the dynamic spinodal in mean-field dynamics,
see Sec.~\ref{secbind}.

In Fig.~{\ref{fig8O}}(a) we show the temperature dependence of
$H_{1/2}$ for $L$$=$$24$.
The MC values lie between the estimates from MFD1 and MFD2.
Again we considered the standard Metropolis dynamic with
spin updates at randomly chosen sites.
The MC simulations in this case were accelerated
by using the method of absorbing Markov chains.
This new MC method \cite{MAN} generalizes the $n$-fold way algorithm
\cite{NFOLD,NOVOTNY_CIP} and gives large CPU-time savings for
low temperatures without changing the underlying dynamics.

Analytic estimates for $H_{\rm DSP}$ at low temperatures can be obtained as
follows. For sufficiently low temperatures there exists
a field $2$$<$$|H|$$<$$4$ such that
a single overturned spin is a supercritical
droplet \cite{NS,SCH,MOS,Sco1,KO,Sco2}.
We define $\tau_1$ as the average time before
a single overturned spin appears.
Further, we define $\tau_2$ as the average first-passage time
from the state with a single overturned spin to
the absorbing state with magnetization $m_{\rm cut}$$=$$0$. For
$2$$<$$|H|$$<4$ and low temperatures,
the processes that determine $\tau_2$ are deterministic, and for
both MFD1 and MFD2, $\tau_2$ is of order unity.
Using the free-energy difference between the
state with a single overturned spin and the metastable state with no
overturned spins, $2 \beta |H| -8\beta+\ln N$,
one can obtain $\tau_1$ for MFD1 and MFD2 from Eqs.~(\ref{efrs6})
and (\ref{efrs9}), respectively.
We get $\tau_1\propto\exp(8\beta-2 \beta |H|- \ln N)$ for MFD1 and
$\tau_1\propto\exp(8\beta-2 \beta |H|-2\ln N)$ for MFD2.
Note that we here give $\tau_1$ in units of MCSS and that
$N$$=$$L^2$.  The waiting time
$\tau_1$ can be either large compared to $\tau_2$,
corresponding to stochastic decay, or small,
corresponding to deterministic decay.
Therefore, the estimate for $H_{\rm DSP}$ can be obtained by setting
$\tau_1$ to be of the same order of magnitude as $\tau_2$,
which leads to
\begin{mathletters}
\label{edspH}
\begin{eqnarray}
\label{edspHa}
4-H_{\rm DSP} & = & T(\ln L + c_1) \quad  {\text{~~for MFD1}}\\
\label{edspHb}
4-H_{\rm DSP} & = &  T(2\ln L + c_2)  \quad  {\text{for MFD2}}~,
\end{eqnarray}
where $c_1$ and $c_2$ are non-universal constants. For MC simulations,
$\tau_1\propto\exp(8\beta-2 \beta |H|- \ln N)$ as for MFD1.
Simulations in which the update sites are chosen sequentially give $\tau_2$
on the order of unity, so that $H_{\rm DSP}$ is given by Eq.~(\ref{edspHa})
for these microscopic dynamics as well.
When updates are performed at randomly chosen sites, however,
$\tau_2$ increases with $L$.
The radial growth velocity $v$ is independent of $L$
and $H$ for $2$$<$$|H|$$<4$ and $T$$\ll$$1$, so that
$v\tau_2=O(L)$, and $\tau_1=\tau_2$ then gives
\begin{equation}
\label{edspHc}
4-H_{\rm DSP}=T \left( \frac{3}{2} \ln L + c_3 \right)
\;.
\end{equation}
\end{mathletters}
In Fig.~{\ref{fig8O}} we compare these analytic low-temperature estimates
for $H_{\rm DSP}$ with $H_{1/2}$ as obtained both from our mean-field
dynamics and from microscopic MC simulations.
Excellent agreement for the linear $T$ dependence is demonstrated in
Fig.~{\ref{fig8O}}(a) and for the logarithmic $L$ dependence in
Fig.~{\ref{fig8O}}(b).

In Fig.~\ref{fig9O} we show the field dependence of the slope
$\Lambda$ for $L$$=$$24$
as a function of $|H|$ for $T$$=$$1.0$, 1.1 and 1.2 in the SD region.
(Note that $T_{\rm c}$$=$$2.269$$\cdots$.)
We observe that, as expected,  the size of the asymptotic SD region
increases significantly as $T$ is lowered.
The size of the MD subregion, which is not shown,
is very compressed for these low temperatures,
since the entropic factor is much less important in $F(m)$. From
Fig.~\ref{fig9O} we confirm that $b$$+$$c$$=$$2.0(1)$ for MFD1 and
$b$$+$$c$$=$$3.7(2)$ for MFD2.
To obtain these estimates of $b$$+$$c$ and $\beta\Xi$, we performed
least-square fits to Eq.~(\ref{efrs28}) considering only data points
in the asymptotic SD region in the following way.
We first discarded data in the CE region.
We then estimated the slopes and the intercepts
as we successively removed data points
from the strong-field end of the $|H|$ interval.
Since the asymptotic region for Eq.~(\ref{efrs28}) corresponds to
$H_{\rm THSP}$$<$$|H|$$\ll$$1$,
our estimates for the slopes and the intercepts
approached constant values with small fluctuations
as we removed data points.
Due to the uncertainty in $S(E,M)$ it is not easy to obtain
a systematic error analysis.
We therefore estimated the errors from the fluctuations of the estimates
in the asymptotic region.
The resulting estimates for $\beta\Xi$ from MFD1 are slightly different
from those of MFD2, as is shown in Fig.~\ref{fig9O}.
The final error bars in our estimates for $\beta\Xi$ include this effect.
Our estimates for $\beta\Xi$, 4.95(2), 4.00(1), and 3.20(2) for
$T$$=$$1.0$, 1.1, and 1.2, respectively,
are within $1 \%$ of the exact values
\cite{ZA,RW} 4.942, 4.004, and 3.217, respectively.

Since the nucleation rate for a single droplet is independent of the
system size, the average metastable lifetime in the SD region should be
proportional to $L^{-2}$, as indicated by Eq.~(\ref{efrs25}).
We checked this result at $T$$=$$1.0$ and
$|H|^{-1}$$=$$2.47$ by fitting the lifetime to the form
$\langle \tau \rangle$$\propto$$L^{- \alpha}$
for $L$=12, 14, 16, 20, and 24
(in the SD region for all the values of $L$ used). From this we obtained
$\alpha$$=$$2.08(3)$, in reasonable agreement with the theoretical result.
At higher temperatures the agreement is less convincing, so that
at 0.8$T_{\rm c}$ we found $\alpha$$\approx$3 for the same values of $L$.
However, we believe this is a finite-size effect which becomes more
pronounced at higher temperatures.

It is also possible to construct mean-field dynamics that
interpolate smoothly between MFD1 and MFD2.  In analogy to
Eqs.~(\ref{efrs6}) and (\ref{efrs9}) the transition probabilities
of such a dynamic can be written as
\begin{mathletters}
\label{efrs_gamma}
\begin{eqnarray}
\label{efrsGa}
W_\gamma(n,n+1) & = &
A \left(1-{n \over N}\right)^\gamma
\exp \left[\min\left\{0,\left(F(n)-F(n+1) +
\gamma\left(S(n)-S(n+1)\right)\right)\right\}\right]\\
\label{efrsGb}
W_\gamma(n,n-1) & = &
A \left({n \over N}\right)^\gamma
\exp\left[\min\left\{0,\left(F(n)-F(n-1) +
\gamma\left(S(n)-S(n-1)\right)\right)\right\}\right]\\
\label{efrsGc}
W_\gamma(n,n) & = & 1 - W_\gamma(n,n+1) - W_\gamma(n,n-1)
\;.
\end{eqnarray}
\end{mathletters}
The positive constant $A$ only needs to fulfill the requirement that
$W_\gamma(n,n) \! \ge \! 0$ for all $n$,
and it is otherwise unimportant since it only redefines
the overall timescale of the process. In principle, the only restriction on
$\gamma$ should be $\gamma \! \ge \! 0$.
{For} $\gamma$$=$$1$ and $A$=1 the process reduces to MFD1, while for
$\gamma$$=$$0$ and $A$=1/2 it reduces to MFD2.  Since both the prefactors
$b$$+$$c$ in the SD
region and the $L$ dependence for the low-temperature strong-field
behavior of $H_{\rm DSP}$ are different for MFD1 and MFD2, it is expected
that they both change continuously with $\gamma$.  Thus it may be
possible to tailor a mean-field dynamic to give the desired
prefactor and $H_{\rm DSP}$ dependence to match a particular microscopic
dynamic. For MC with sites selected sequentially, MFD1 has the desired
values of these two quantities.  If $\gamma$$=$$1$$/$$2$, then the
$T$ dependence of $H_{\rm DSP}$ will be given by Eq.~(\ref{edspHc}),
which agrees with MC with randomly selected sites.  The mean-field
dynamic with $\gamma$$=$$1$$/$$2$ might also be expected to give
a prefactor closer to the value for MC with randomly selected sites,
$b$$+$$c$$=$$3$ \cite{RTMS}, since this lies between the values we have
obtained for MFD1 and MFD2.

In Fig.~\ref{figdisc} we show $T \Lambda$ for the very low
temperatures $T$$=$$0.4$, 0.2, and 0.1.
Discrete-droplet results for low $T$ \cite{NS,SCH,MOS,Sco1,KO,Sco2}
are shown as a set of parabolic arcs given by Eq.~(\ref{efrs41}).
For $T$$=$$0.4$ we observe clear oscillatory behavior with $|H|$,
even though quantitative agreement with
the discrete-droplet limit has not set in yet.
For $T$$=$$0.2$ and $0.1$ we observe increasingly good quantitative
agreement between the MFD1 results and the discrete-droplet limit.
For $T$$=$$0.4$, $\langle\tau\rangle$ is on the order of
$10^7$ and $10^{26}$ for $|H|$$=$$1.0$ and $|H|$$=$$0.3$,
respectively, as obtained from MFD1 with $m_{\rm cut}$$=$$0$.
The corresponding numbers for $T$$=$$0.2$\ $(0.1)$ are
$10^{18}$\ $(10^{39})$ and $10^{59}$\ $(10^{123})$.

MC results by one of us \cite{MAN} provide corroboration
for the MFD1 results.
This oscillatory behavior with $|H|$ is due to the
corrugation of the free energy near the saddle point, illustrated in
Fig.~\ref{figfm24}(b).
This corrugation is due to the discreteness of the lattice.

\section{Relations to Earlier Work}
\label{secbind}

The relationships of the detailed shape of $F(m)$ to two-phase equilibria
and nucleation barriers in finite systems have previously been studied by
Binder and coworkers \cite{KB,FB,KASKI}. Although their work covers a
wide range of temperatures below $T_{\rm c}$, the bulk of their
analysis is concerned with the critical region, in which
the correlation lengths become comparable to the other length scales,
$L$, $R_0$, and $R_{\rm c}$ (see also Ref.~\cite{MON87}).
Since $F(m)$ can be viewed as
a potential for the stochastic processes that define our macroscopic
dynamics, it might have
been expected that those studies also would have revealed two $L$-dependent
spinodals with the same scaling behaviors as $H_{\rm THSP}$ and
$H_{\rm DSP}$. However, they instead reported a single spinodal at a
field proportional to $L^{-1}$ \cite{FB}.

To clarify the relation between our results and those of these earlier
studies, we follow Furukawa and Binder \cite{FB} by calculating
\begin{equation}
\label{eqFB}
h(m) = (\beta L^d)^{-1} \frac{{\rm d}F_0(m)}{{\rm d}m} \;.
\end{equation}
This quantity has the dimension of a field and is the equivalent, for
an Ising model in the fixed-$m$ ensemble, of the expectation value of the
chemical potential for a lattice gas in the fixed-density ensemble \cite{FB}.
In Fig.~\ref{figFB} we show $h(m)$ for $L$=24, 32, 64, and 96 at
0.8$T_{\rm c}$ \cite{NOTE2}.
(For $L$=96 the simulation was only performed for $|m|$$>$0.63.) From
Eqs.~(\ref{eqFdef}) and~(\ref{eqFB}) it is seen that
\begin{equation}
\label{eqFB2}
\frac{{\rm d}F(m)}{{\rm d}m} = \beta L^{d} \left( h(m) - H \right) \;.
\end{equation}
Thus the extrema of the field-dependent
$F(m)$ occur where the applied field $H$ equals $h(m)$. For a negative
$H$ between zero and the minimum of $h(m)$, there are three such
extrema: the stable minimum at $m$$\approx$$-$$m_{\rm eq}(T)$, the metastable
minimum at $m$$\approx$$+$$m_{\rm eq}(T)$, and an unstable maximum at
some $m$ between 0 and $m_{\rm sp}$, the magnetization corresponding
to the minimum of $h(m)$.
If dynamically relevant information for nonzero $H$
is to be deducible from the functional form of $F(m)$, this must mean that
the spatial configuration characteristic of the nonequilibrium saddle point
of the relaxing system is well approximated by the {\it equilibrium}
configuration corresponding to the same value of $m$ in the fixed-$m$
ensemble. It is not unreasonable to expect that this should hold,
at least for sufficiently large systems that the corresponding applied
$H$ is weak.

Following the reasoning outlined above, we start by considering the
thermodynamic spinodal. The droplet-theoretical arguments in
Sec.~\ref{secdt} indicate that the THSP should be located in the field range
where $R_{\rm c}$ becomes on the order of $L$, so that the droplet
must compete with slab configurations about being the ``true'' saddle point.
It has been shown by Leung and Zia \cite{LZ} that for
Ising models with periodic boundary conditions in the fixed-$m$
ensemble, a first-order phase transition between the equilibrium droplet
and slab configurations occurs at a temperature dependent magnetization,
$m_{\rm c}(T)$. For $|m|$$<$$m_{\rm c}$ the equilibrium configuration is
a slab. The volume fraction corresponding to $m_{\rm c}(T)$
in the limit $L \! \rightarrow \! \infty$,
which was obtained in Ref.~\cite{LZ}, can be written as \cite{NOTE3}
\begin{equation}
\label{eq:hthsp2a}
\phi_{\rm c}(T) = \left( \frac{2 \sigma_0(T)}{d}\right)^{\frac{d}{d-1}}
               \left( \frac{d \! - \! 1}{2 m_{\rm eq}(T)} \right)
               \Xi(T)^{-{\frac{1}{d-1}}} \;.
\end{equation}
The vertical, dashed lines in Fig.~\ref{figFB} mark
$m_{\rm c}(0.8T_{\rm c})$.
If one ignores entropy effects due to the center-of-mass
positions of the droplet and slab,
as well as capillary waves on their surfaces (which are responsible for
the $|H|^b$ power-law prefactor in the nucleation rate \cite{GNW}),
the finite-size displacement of $m_{\rm c}$ is proportional to
$L^{-1}$ \cite{LZ}. The vertical arrows in Fig.~\ref{figFB}
indicate the corresponding values of $m_{\rm c}$ for $L$=24, 32, and 64,
as obtained from Fig.~3 of Ref.~\cite{LZ}.
These values monotonically approach the
infinite-$L$ limit for $m_{\rm c}$. From Fig.~\ref{figFB}(a) it appears that
the shape of $h(m)$ evolves towards a step
discontinuity as $L$ increases. From the data collapse in the plots
of $L h(m)$ in Fig.~\ref{figFB}(b),
it is seen that the magnitude of this discontinuity
is proportional to $L^{-1}$ and so corresponds to a finite discontinuity
in a first derivative of the free energy per unit interface area,
$L^{-(d-1)}F_0(m)$, in the
fixed-$m$ ensemble. This is consistent with the identification
of the droplet-to-slab transformation as
a first-order phase transition \cite{LZ}.

Using $\phi_{\rm c}(T)$ from Eq.~(\ref{eq:hthsp2a}) in Eq.~(\ref{eq:hthsp})
for $H_{\rm THSP}(\phi)$, we obtain an estimate for $H_{\rm THSP}$
as the field at which the free energies of the critical droplet and
a system-spanning slab are degenerate:
\begin{equation}
\label{eq:hthsp2b}
H_{\rm THSP} = H_{\rm THSP}(\phi_{\rm c})
= \frac{1}{L} \left( \frac{d \ \Xi(T) }{2 \sigma_0(T)} \right)^{\frac{1}{d-1}}
\;.
\end{equation}
The horizontal, dashed line in Fig.~\ref{figFB}(b) represents this estimate
of $L\>H_{\rm THSP}$. It corresponds excellently to the magnitude of the
incipient step discontinuity in $L\>h(m)$.
The $L$ dependence of $H_{\rm THSP}$ at 0.8$T{\rm c}$ is shown in
Fig.~\ref{figH}, together with the results of MFD1 for $L$=32 and~64.
The MFD1 points were estimated as the smallest fields for which
$\Lambda(T,H)$=$\beta \Xi(T)$ (see Fig.~\ref{fig3O}).
The agreement is good, and we believe the small
discrepancy is a finite-size effect.

The narrow minimum in $h(m)$, which occurs at an $L$-dependent magnetization
$m_{\rm sp}(L)$, slightly smaller than the equilibrium magnetization
$m_{\rm eq}$, signifies the inflection point in $F_0(m)$.
As pointed out by Furukawa and Binder \cite{FB},
this magnetization also corresponds to a change in
the equilibrium configuration: closer to
$m_{\rm eq}$ the system is uniform, whereas closer to $m_{\rm c}$ a
single droplet of the opposite magnetization is precipitated. The volume
fraction occupied by this droplet is given by the lever rule.
As seen from Fig.~\ref{figFB}(b), the minimum value of $h(m)$, $h(m_{\rm sp})$,
does not vanish as $L^{-1}$.

It is tempting to identify the disappearance of the single-droplet
saddle point, which occurs at $h(m_{\rm sp})$,
with the dynamic spinodal, $H_{\rm DSP}$. For all the system sizes studied
here, we find that $|h(m_{\rm sp})|$ indeed lies close to other estimates
for $H_{\rm DSP}$, such as $H_{1/2}$ obtained both from MFD1 and from
microscopic MC simulations. These comparisons are illustrated in
Fig.~\ref{figH}.
Obviously, the range of $L$ used in the present study
is too narrow to obtain the scaling relation with
any degree of certainty. However, the values of
$|h(m_{\rm sp})|$ at different $L$ also
agree well with the estimate for $H_{\rm DSP}$ obtained
in Ref.~\cite{RTMS} by
fitting the proportionality constant in the relation
$L$$\propto$$R_0$ with $R_0$ given by
Eq.~(\ref{efrs29}) to MC data for $L$ between~64 and~720.
We find this noteworthy, considering that this analytical
expression for $H_{\rm DSP}$ is based on explicitly dynamical arguments,
whereas $|h(m_{\rm sp})|$ is obtained from a strictly equilibrium calculation.

To account for this observation, we suggest that $m_{\rm sp}$ corresponds
to the volume fraction at which entropy effects make the free energy of
a configuration consisting of two droplets lower than that of the
single-droplet configuration \cite{FB}.
We assume that the single droplet is replaced by two identical
droplets, each with half the volume of the original droplet.
We neglect corrections to the
droplet free energy, including those
arising from surface excitations, which correspond
to the power-law prefactor $|H|^b$
in Eq.~(\ref{efrs24}).
We only consider the entropy contributions due to the droplets'
center-of-mass positions, and we neglect excluded-volume effects.
This gives the approximate
zero-field free energies
\begin{mathletters}
\begin{equation}
F_0^{(1)}(\phi) \approx F_0(m_{\rm eq}(T)) + L^{d-1} d \beta
\left( \frac{2 m_{\rm eq}(T) \phi }{d \! - \! 1}\right)^{\frac{d-1}{d}}
\Xi(T)^{\frac{1}{d}} - d \ln L
\label{eq:f1d}
\end{equation}
for the single-droplet configuration and
\begin{equation}
F_0^{(2)}(\phi) \approx F_0(m_{\rm eq}(T))
+ 2^{\frac{d-1}{d}} L^{d-1} d \beta
\left( \frac{2 m_{\rm eq}(T) \phi }{d \! - \! 1}\right)^{\frac{d-1}{d}}
\Xi(T)^{\frac{1}{d}} - 2 d \ln L
\label{eq:f2d}
\end{equation}
for the two-droplet configuration.
\end{mathletters}
Equating $F_0^{(1)}$ and $F_0^{(2)}$ we find
\begin{equation}
\label{eq:phisp}
\phi_{\rm sp} \approx
\frac {d \! - \! 1}{2 m_{\rm eq}(T) \Xi(T)^{\frac{1}{d-1}}L^d }
\left( \frac{T \ln L}{2^{\frac{d-1}{d}} \! - \! 1} \right)^{\frac{d}{d-1}}
\;.
\end{equation}
Inserting $\phi_{\rm sp}$ into the single-droplet
approximation for $|h(\phi)|$ obtained from $F^{(1)}$
through Eq.~(\ref{eqFB}),
which is identical to the expression for $H_{\rm THSP}(\phi)$ given in
Eq.~(\ref{eq:hthsp}), we obtain
\begin{equation}
\label{eq:hphisp}
|h(\phi_{\rm sp})| \approx
\left( \frac{\left(2^{\frac{d-1}{d}} \! - \! 1\right) \beta \Xi(T)}{\ln L}
\right)^{\frac{1}{d-1}}
\;.
\end{equation}
Comparing this result with Eq.~(\ref{efrs30}) for $H_{\rm DSP}$ we note that,
except for a $d$-dependent
numerical constant of order unity, it has the same asymptotic
dependences on $L$ and $T$ as $H_{\rm DSP}$.
In particular, $|h(\phi_{\rm sp})|$ does {\it not} vanish as $L^{-1}$,
but rather much more slowly as $(\ln L)^{-\frac{1}{d-1}}$.
As noted previously and illustrated in
Fig.~\ref{figH}, corrections to this asymptotic behavior for $H_{\rm DSP}$,
due to the power-law prefactors in the nucleation rate, are very substantial.
Since analogous corrections were ignored in the approximate derivation of
$|h(\phi_{\rm sp})|$ given here, we expect similarly large corrections
to apply to it. This is in agreement with the numerical results shown in
Figs.~\ref{figFB} and~\ref{figH}.
We further note that the precise value of the numerical coefficient in
our approximate expression for the asymptotic value of $|h(\phi_{\rm sp})|$
is the result of our choice to consider only a separation of the single
droplet into two {\it equal} droplets. A more careful calculation ought
therefore to give a different coefficient. However, the factor
$(d \! + \! 1)$ in the denominator of $H_{\rm DSP}$ results specifically from
the simultaneous nucleation and growth processes which give rise to
the expression for the lifetime in the MD region, Eq.~(\ref{efrs27}). It is
therefore unlikely that further improvement of the equilibrium calculation
should yield the same factor in $|h(\phi_{\rm sp})|$.

{For} $|H|$$>$$|h(m_{\rm sp})|$ the bulk free energy displays
no saddle point, but its derivative with respect to $m$ has a minimum
at $m_{\rm sp}$. As a consequence, mean-field dynamics should show a
slowing-down near $m_{\rm sp}$, as observed in the present study. However,
$F(m)$ does not contain information about the complicated multi-droplet
configurations that dominate the dynamics in this region of relatively
strong fields. Consequently we expect the agreement between microscopic
and mean-field dynamics to be only qualitative in the MD region.

\section{Discussion}
\label{secdis}

In this paper we have introduced a class of macroscopic mean-field
dyamics, and have studied in detail two members of this class which
we call MFD1 and MFD2.
We have demonstrated that these macroscopic mean-field dynamics
replicate many of the qualitative and
quantitative features of the relaxation behavior
of the metastable phase of the two-dimensional nearest-neighbor
Ising model.
As a function of the external field $H$, four distinct regions of
relaxation were observed in agreement with recent microscopic
MC studies \cite{RG94,RTMS}.
In the single-droplet region at low temperatures,
the leading exponential term in the relaxation time
$\langle\tau\rangle$ was obtained to within $1\%$ of the exact value.
We also obtained temperature independent estimates for the prefactor
exponents $b$$+$$c$ for the two mean-field dynamics.
At very low temperatures we observed an oscillatory behavior in
$\Lambda$$($$H$$)$$\equiv$${\rm d}$$\ln\langle \tau \rangle/ {\rm d} |H|^{-1}$
with respect to $|H|$,
in agreement with discrete-droplet theory.
In the low-temperature limit, simple theoretical estimates of
the dynamic spinodal field $H_{\rm DSP}$,
in terms of the temperature and the system size, were obtained
for various dynamics.
Our numerical studies provide excellent agreement with these predictions.

The mean-field dynamics are constructed with only the following conditions:
(1) locality in the value of the order parameter $m$, and
(2) the correct equilibrium distribution obtained
from the order parameter of the microscopic model.
These two conditions constitute the minimum requirements for any
local dynamic.
The reasonable results obtained from the macroscopic mean-field dynamics
may be somewhat surprising in view of the
fact that no microscopic information is directly included.
However, we believe the relative success of our dynamics
becomes more understandable if one views them as approximations
to the macroscopic dynamic which can formally be constructed
by using a projection-operator formalism to project the microscopic
dynamic onto a master equation for the one-dimensional order parameter
distribution, as discussed in Appendix~\ref{app_pro_op}.
The dynamically relevant information which is retained in $F(m)$
following this projection correctly describes the droplet configurations
that provide the rate-determining steps in the decay process. In particular,
these are the droplet and slab configurations that are important near the
thermodynamic spinodal, the single-droplet configuration characteristic of
the SD region, and the breakdown of the single-droplet configuration
into a uniform ``gas'' of microscopic fluctuations that takes place near the
dynamic spinodal.
These aspects were discussed in light of earlier work in
Sec.~\ref{secbind}.
As a consequence, the mean-field dynamics produce excellent numerical
estimates for both the thermodynamic and the dynamic spinodal fields.

Comparison between the two proposed dynamics,
MFD1 and MFD2, provides some insight
into the dependence of the lifetime $\tau$ on the detailed dynamics.
Most of the characteristic behavior of
$\Lambda$
predicted by both continuous- and discrete-droplet theory
\cite{RTMS,NS} is expected to hold for different local dynamics.
However the influence of the detailed dynamic is reflected in
the prefactor exponent $b$$+$$c$.
The difference between  $\Lambda(H)$ as obtained from MFD1 and MFD2
in the deterministic region suggests that
$\Lambda(H)$ in this region depends strongly on the particular dynamic.

Generally, when the relaxation time is long,
the system spends more time exploring phase space,
and therefore the dynamic is more strongly subject to the
restricted bulk free energy $F(m)$, as indicated in this study.
When the relaxation time is short,
the details of the particular dynamic are more important than
the bulk free energy. From the field-theoretical point of view,
when the most probable
trajectory from the metastable phase to the unstable saddle point
is sharply defined, the relaxation time is usually large, and
regardless of the details of the particular dynamic, one needs to
consider only the trajectories near the most probable one
in order to study the decay of the metastable phase.
When the probability distribution over trajectories
is not very sharp, the lifetimes are usually
short and the details of the particular dynamic play important roles.

The biggest advantage of macroscopic mean-field dynamics is that
they can provide data for low temperatures
and weak fields (in the single-droplet region),
for which
the average lifetime of the metastable phase is too long to be measured
with standard MC algorithms.
They also provide data for arbitrary temperatures and fields,
allowing one to obtain accurate estimates of derivatives.
However, the system sizes for which mean-field dynamics
can be applied are rather limited by computational constraints.

In summary, we have presented a method to
study the relevance of the equilibrium properties
of a model to the dynamical relaxation of metastable phases.
The macroscopic dynamics
are designed using only the minimal requirements of
locality in the relevant order parameter
and the correct equilibrium free energy projected on that
order parameter.
Extensive applications to the two-dimensional
nearest-neighbor Ising ferromagnet
on a square lattice provides convincing evidence that the characteristic
behavior of the dynamical relaxation of the metastable phases is largely
determined by the restricted bulk free energy $F(m)$.
We believe that our approach can benefit studies
of relaxation phenomena for other systems as well.

\acknowledgements
We would like to thank K.~Binder, M.~Grant, J.~M.\ Kosterlitz, and
J.~Vi{\~{n}}als for useful discussions,
and B.~M.\ Gorman, C.~C.~A.\ G{\"u}nther, R.~A.\ Ramos, H.~L.\ Richards,
and S.~W.\ Sides for helpful comments on the manuscript.
This work was supported in part by Florida State University
through the Supercomputer Computations Research Institute
(U. S.  Department of Energy Contract No.~DE-FC05-85ER25000).
PAR is also supported by National Science Foundation
Grant No.~DMR-9315969 and by the Florida State University
Center fo Materials Research and Technology (MARTECH).

\newpage
\appendix

\section{Relation of our Macroscopic Magnetization Dynamics to the
Microscopic Spin Dynamic}
\label{app_pro_op}

In this Appendix we briefly consider the formal relationship between our
macroscopic magnetization dynamics, MFD1 and MFD2,
and the underlying microscopic dynamic represented by the single-spin-flip
Metropolis algorithm with updates at randomly selected sites.
The formal framework for the discussion is the
Nakajima-Zwanzig \cite{NAKA58,ZWAN60}
projection-operator formalism for the master equation,
which is equivalent to Mori's \cite{MORI65}
projection-operator formalism for the equations of motion of observables
\cite{GRAB77C,GRAB82}. We adapt the standard discussion (see, {\it e.g.},
Refs.~\cite{GRAB82,NORD75}), which considers
a deterministic microscopic dynamic governed by a quantum-mechanical
or classical
Hamiltonian, to the case where the microscopic dynamic is a discrete-time
Markov process (possibly derived from a deterministic dynamic at an even
more microscopic level).

For an $N$-site kinetic
Ising model, the microscopic probability density at time $k$
is a $2^N$-dimensional column vector $\vec{\rho}(k)$ which evolves in time
according to the equation
\begin{equation}
\label{eqA1}
\vec{\rho}(k \! + \! 1) = {\cal W} \vec{\rho}(k) \;,
\end{equation}
where $\cal W$ is the matrix of microscopic transition probabilities.
The probability distribution over the ``relevant'' macroscopic variables
at time $k$, $\vec{X}(k)$, is obtained
from $\vec{\rho}(k)$ through the action of a projection operator $\cal P$,
\begin{equation}
\label{eqA2}
\vec{X}(k) = {\cal P} \vec{\rho}(k) \;.
\end{equation}
Although the dimension of $\vec{X}(k)$ is $2^N$,
if the only macroscopic variable considered is the magnetization, the
dimension of the ``relevant'' space in which $\vec{X}(k)$ has
nonzero components is $N$+1.
By using Eqs.~(\ref{eqA1}) and~(\ref{eqA2}), one can write the equation of
motion for $\vec{X}(k)$ as
\begin{eqnarray}
\label{eqA3}
\vec{X}(k \! + \! 1) &=& {\cal P} {\cal W} \vec{\rho}(k)
              =  {\cal P} {\cal W} ({\cal P} \! + \! {\cal Q} ) \vec{\rho}(k)
                 \nonumber\\
             &=& {\cal P} {\cal W} \vec{X}(k)
               + {\cal P} {\cal W} \left[ {\cal Q} \vec{\rho}(k) \right]
\;,
\end{eqnarray}
where $\cal Q$=$\bf 1$$-$$\cal P$ is the projection operator onto the
``irrelevant'' orthogonal complement to the relevant space.
The first term in the second line of Eq.~(\ref{eqA3}) corresponds to a
Markov process for $\vec{X}(k)$, whereas the second term contains non-Markov
contributions which can be formally evaluated as follows.

Operating on Eq.~(\ref{eqA1}) with $\cal Q$ one obtains
\begin{eqnarray}
\label{eqA4}
{\cal Q} \vec{\rho}(k) &=& {\cal Q} {\cal W} \vec{\rho}(k \! - \! 1)
  =  {\cal Q} {\cal W} ({\cal P} \! + \! {\cal Q} ) \vec{\rho}(k \! - \! 1)
                 \nonumber\\
 &=& {\cal Q} {\cal W} \vec{X}(k \! - \! 1)
   + {\cal Q} {\cal W} \left[ {\cal Q} \vec{\rho}(k \! - \! 1) \right]
\;,
\end{eqnarray}
which is inserted in Eq.~(\ref{eqA3}). Iterating this procedure a total of
$k$ times, one obtains the final result:
\begin{equation}
\label{eqA5}
\vec{X}(k \! + \! 1) = {\cal P} {\cal W} \vec{X}(k)
+ {\cal P} {\cal W} \sum_{l=1}^k \left[ {\cal Q} {\cal W} \right]^l
  \vec{X}(k \! - \! l)
+ {\cal P} {\cal W} \left[ {\cal Q} {\cal W} \right]^k
  {\cal Q} \vec{\rho}(0) \;.
\end{equation}
The first term on the right-hand side corresponds to a Markov process in
the relevant subspace.
The non-Markovian second and third terms represent memory about
the relevant variables at earlier times,
propagated through the irrelevant subspace, and specific information about
the initial state of the irrelevant variables, respectively \cite{NORD75}.
The third term can usually be ignored, at least after a short initial period.

In standard applications of projection-operator techniques, the quality of the
resulting approximation depends on the choice of the
relevant macroscopic variables. The approach is most useful whenever there is
a large separation between ``fast'' and ``slow'' timescales, and one usually
attempts to include all the slow variables in the relevant subspace.
This ensures that only variables with short correlation times contribute to
the memory effects, which can then often be ignored or approximated by a
rapidly decaying function. The (most obvious) slow variables are determined
by macroscopic conservation laws or by spontaneously broken symmetries
in the corresponding isolated system \cite{GRAB82}.

In the present work we have considered the stochastic time evolution of the
magnetization (our relevant macroscopic
variable) as a Markov process defined by
the transition probability matrices ${\bf W}_1$ (for MFD1) or ${\bf W}_2$
(for MFD2). In doing so we have performed a Markov approximation equivalent to
ignoring the memory effects and using ${\bf W}_1$ and ${\bf W}_2$
as approximations for the matrix obtained by contracting
${\cal P} {\cal W}$ in Eq.~(\ref{eqA5}),
so that its dimension becomes $N+1$.
The macroscopic slowness of the magnetization is related to the
spontaneously broken symmetry
between the two ferromagnetic phases for $H$=0 below $T_{\rm c}$.
By virtue of energy conservation in the corresponding closed system,
the other obvious slow macroscopic variable is the
total energy. In relegating it to the irrelevant subspace, mainly for
computational convenience, we have most likely ignored non-negligible
memory effects. By considering only a single relevant variable we also
have excluded nonlinear interactions between
macroscopic variables \cite{GRAB82}.

The satisfactory
agreement between our approximate macroscopic Markovian dynamics and the
MC simulations of the full microscopic dynamic indicates that the
approximations made in the present work are quite reasonable.
Nevertheless, the above discussion
indicates that by including the total energy as a second relevant
macroscopic variable
in our Markovian mean-field dynamics, we could reduce the importance of
the neglected memory effects and allow for nonlinear interactions between
relevant variables.
We believe further significant improvement of the agreement between the
approximate, macroscopic dynamics and the underlying microscopic
dynamic could be achieved in this way.

\newpage
\section{Absorbing Markov Chains}
\label{app_amc}

In this Appendix we briefly discuss the application
of absorbing Markov chains \cite{AMC} to obtain the expectation value
and the variance of the first-passage time
for escape from the metastable phase.
For a given transition probability matrix $W(n,n^{\prime})$,
simple matrix calculations provide
the expectation value and the standard deviation
of the first-passage time from one state to another.

Let us consider a random walker which moves between states
$0 \le n,n^{\prime} \le N$ with transition probabilities
$W(n,n^{\prime})$.
Starting from an arbitrary initial state
$ \vec X(0)$,
the probability density after $k$ time steps, $\vec X(k)$,
is
\begin{equation}
\label{efrs10}
 \vec X(k)={\bf W}^k \vec X(0)
 \;.
\end{equation}
Without absorbing states, the probability of a
Markov chain is conserved, that is,
\begin{equation}
\label{efrs11}
\sum_{n^{\prime}} W(n,n^{\prime}) = 1
\end{equation}
for all $n$.
After $k$ time steps the walker is still in some state, that is
\begin{equation}
\label{efrs12}
\vec e^{\rm T} {\bf W}^k \vec X(0) = 1
 \;,
\end{equation}
where ${\vec e}^{\rm T} = (1,\cdots,1,\cdots,1)$ and
the superscript ${\rm T}$ denotes the transpose.

Now we place absorbing states at $i \ge n_{\rm cut}$.
Once an absorbing state is reached, the walker is
absorbed and the Markov chain terminates.
Let ${\bf T}$ be the $n_{\rm cut} \times n_{\rm cut}$
submatrix of ${\bf W}$ that contains the transition
probabilities between the $n_{\rm cut}$ transient states.
The analog of Eq.~(\ref{efrs11}) is not satisfied for ${\bf T}$.
The probability that the walker is absorbed at time $k$ is
$\vec e^{\rm T} ({\bf T}^{k-1}- {\bf T}^k) \vec X(0)$.
The average first-passage time to the absorbing states is
\begin{equation}
\label{efrs13}
\langle\tau\rangle = \sum _{k=1}^\infty
 \vec e^{\rm T}k({\bf T}^{k-1}-{\bf T}^k) \vec X(0)
 =  \vec e^{\rm T} {{\bf N}} \vec X(0)
 \;.
\end{equation}
The fundamental matrix is defined by ${\bf N} = ({\bf I-T})^{-1}$,
where ${\bf I}$ is the identity matrix.
Similarly, the second moment of the first-passage time
can be obtained from \cite{AMC}
\begin{equation}
\label{efrs14}
\langle\tau ^2\rangle = \vec e^{\rm T}(2{\bf N}^2- {\bf N}) \vec X(0)
 \;.
\end{equation}

In this work, $\tau$ was divided by the total number of sites $N$,
so that all times are given in units of MCSS.

\newpage

\newpage

\begin{figure}
\caption[]{
(a) The zero-field bulk restricted free energy, $F_0(m)$,
of the nearest-neighbor Ising ferromagnet on a square lattice
at $T$$=$$0.8T_{\rm c}$$=$$1.8153$$\dots$ for $L$$=$$24$.
(b) Same as (a), but at $T$$=$$0.2$$\approx$$0.0881T_{\rm c}$.
Notice the difference in the scales along the
$y$-axis in (a) and (b).
The sawtooth-like behavior of $F_0(m)$ at $T$$=$$0.2$
is due to the discreteness of the lattice.
The vertical arrows marked $m_{\rm c}$
indicate the exactly known magnetization
for which the most likely  configuration with the given magnetization
changes from a slab (for $|m| < m_{\rm c}$) to a droplet
(for $|m| > m_{\rm c}$)
in an infinite system \protect\cite{LZ}.
}
\label{figfm24}
\end{figure}

\begin{figure}
\caption[]{
Schematic plot of
$\Lambda(H)$, defined in Eq.~(\ref{efrs26}), for a two-dimensional
Ising ferromagnet.
The dynamic spinodal field $H_{\rm DSP}$ separates the stochastic and
the deterministic regions.
In the stochastic region ($|H|$$<$$H_{\rm DSP}$),
the relaxation time $\langle\tau\rangle$ is determined by the
formation of a single critical droplet.
Depending on the size of the critical droplet relative to the system
size, the stochastic region is divided into the single-droplet
(SD) and coexistence (CE) subregions.
In the SD region the size of the critical droplet
is smaller than the system size.
The CE region is also characterized by a single nucleating droplet,
but of a size comparable to the system size.
The thermodynamic spinodal field $H_{\rm THSP}$
separates the SD and CE regions.
The deterministic region ($|H|$$>$$H_{\rm DSP}$) is
comprised of the multi-droplet (MD) and strong-field (SF) regions.
These two regions are separated by the mean-field spinodal (MFSP).
Droplet theory \protect\cite{RTMS} predicts that
the intercepts of the two straight lines
are $\beta\Xi$ and $\beta\Xi/3$. Their slopes are related
as shown in the figure if one assumes that the radial growth velocity
of the supercritical droplets is proportional to $H$.
}
\label{fig1O}
\end{figure}

\begin{figure}
\caption[]{
The field dependence of the average
metastable lifetime $\langle\tau\rangle$
for a two-dimensional Ising model at $T$$=$$0.8T_{\rm c}$.
The lifetime $\langle\tau\rangle$ is estimated
as the average first-passage time to $m_{\rm cut}$$=$$0$
from the starting configuration $m$$=$$+1$ with $H$$<$$0$.
The lines correspond to the MFD1 dynamic for $L$$=$$32$ (dashed)
and $64$ (solid).
The results of Metropolis MC simulations with
random site updates for $L$$=$$32$ and $64$
are marked by $+$ and $\times$, with
$10^3$ escapes from the $m$$=$$+$$1$ state
the statistical errors smaller than the symbol size.
}
\label{fig2O}
\end{figure}

\begin{figure}
\caption[]{
(a) The slope $\Lambda(H)$,
obtained from the data in Fig.~\protect\ref{fig2O},
shown as a function of $|H|$.
The thick solid curve is from MFD1 for $L$$=$$64$, and the
symbols are from the standard Metropolis MC simulations with
random site updates for $L$$=$$64$.
The two straight lines are drawn with slopes
$b$$+$$c$$=$$2$ and $(b$$+$$c+2)/3$$=4/3$, using the exact value
\protect\cite{ZA,RW} of $\beta\Xi(0.8T_{\rm c})=0.5062$$\dots$.
The horizontal arrows mark the exact values of $\beta\Xi(0.8T_{\rm c})$
and $\beta\Xi(0.8T_{\rm c})/3$.
The asymptotic SD subregion seems to be
very small for $L$$=$$64$.
The SF subregion ($|H|$$>$$2$),
where $\Lambda(H)$ decreases to zero is not shown.
We define $H_{\rm max}$ and $H_{\rm min}$
as the fields at which $\Lambda$ has a local maximum and minimum,
respectively.
(b) The slope $\Lambda(H)$, shown as a function of $|H|^{-1}$.
The two horizontal lines correspond to the exact values of
$\beta\Xi$ and $\beta\Xi/3$.
The interpretations of the other lines and symbols are the same as in (a).
}
\label{fig3O}
\end{figure}

\begin{figure}
\caption[]{
The field dependence at $T$$=$$0.8$$T_{\rm c}$ of $\Lambda(H)$
for the two different mean-field dynamics, MFD1 and MFD2.
Qualitatively similar behavior is observed for both dynamics.
Using the exact value of $\beta\Xi$,
we estimate in the SD region $b$$+$$c$$\approx$$2.0$ and $3.7$
for MFD1 and MFD2, respectively.
In the MD region, the lifetimes for MFD2 are shorter than for MFD1.
Even though the system size is too small to discuss the asymptotic
behavior in the MD region, the results from MFD1 and MFD2
are not in good agreement with droplet theory in this region.
}
\label{fig5O}
\end{figure}

\begin{figure}
\caption[]{
The field dependence of the relative standard deviation
$r$$=$$\sigma_{\tau} / \langle\tau\rangle$ at $T$$=$$0.8$$T_{\rm c}$.
The curves are from MFD1 for $L$$=$$24,$ $32$, and $64$.
The symbols are from the Metropolis MC simulations with
random site updates for $L$$=$$32$ and $64$.  The MC data are
from $3000$ escapes from the metastable state near $r$$\approx$$1/2$, and
at least $100$ escapes from the metastable state for the other $H$ values.}
\label{fig6O}
\end{figure}

\begin{figure}
\caption[]{
(a) The temperature dependence of the estimate for the
dynamical spinodal field $H_{\rm DSP}$,
given by $H_{1/2}$ for $L$$=$$24$.
Note that the MC simulation results (Metropolis with updates at
randomly selected sites, MR) fall between
the MFD1 and MFD2 results.
The estimates approach $H$$=$$4$ linearly in $T$
with negative slopes, which depend on the system size $L$
and a non-universal constant.
The slopes of the straight lines are 5.7, 4.0, and 3.0.
(b) The asymptotic slopes near the point $(T,H_{1/2})$$=$$(0,4)$
as functions of $L$.
We also show the MC results from the microscopic Metropolis
algorithms with sequential (MS) and random (MR) spin updates.
For comparison we show straight lines with the predicted slopes
(2, 1.5, and 1 from top to bottom),
obtained from the analytic low-temperature estimates for $H_{\rm DSP}$
in Eq.~(\protect\ref{edspH}).
}
\label{fig8O}
\end{figure}

\begin{figure}
\caption[]{
The field dependence of the slope $\Lambda(H)$
for $L$$=$$24$ and $T$$=$$1.0$, $1.1$, and $1.2$.
The size of the asymptotic SD region increases as $T$ is lowered.
We estimate $b$$+$$c=2.0(1)$ and $3.7(2)$ for MFD1 and MFD2, respectively.
The differences between the average values of $\beta\Xi$ for MFD1 and MFD2
and the exact value of $\beta\Xi$ \protect\cite{ZA,RW}
are less than $1 \%$.
Arrows indicate the exact values of $\beta\Xi$ and the lines are from
the fitted asymptotic behavior of $\Lambda(H)$ in the SD region.
}
\label{fig9O}
\end{figure}
%

\begin{figure}
\caption[]{
(a) The field dependence of $T \Lambda(H)$ for MFD1 at $T$$=$$0.4$.
The discrete-droplet \protect\cite{NS,SCH,MOS,Sco1,KO,Sco2}
result is shown as a set of parabolic arcs.
The exact value of $\Xi$ \protect\cite{ZA,RW} is indicated by
the horizontal arrow.
Simulation results for $L$$=$$24$ \protect\cite{MAN} from
the Metropolis algorithm with spin updates at randomly
chosen sites and $10^3$ escapes from the initial state are also shown.
The two vertical arrows indicate the estimates of
$H_{1/2}$ from MC simulation (left) and MFD1 (right).
(b) Same as (a) at $T$$=$$0.2$.
The results from the MC simulations and MFD1
agree quite well with the discrete-droplet
\protect\cite{NS,SCH,MOS,Sco1,KO,Sco2} result.
(c) Same as (a) at $T$$=$$0.1$.
The agreement between the three sets of results is close.
Here the relaxation time $\langle\tau\rangle$ from MFD1
with $m_{\rm cut}$$=$$0$ is on the order of
$10^{39}$ and $10^{123}$ at $|H|$$=$$1.0$ and $0.3$, respectively.
The MC estimate for $\langle\tau\rangle$ is on the order of $10^{31}$
at $|H|$$=$$1.35$.
}
\label{figdisc}
\end{figure}

\begin{figure}
\caption[]{
The quantity $h(m)$, defined in Eq.~(\protect\ref{eqFB}),
shown {\it vs} magnetization $m$
at $T$$=$$0.8$$T_{\rm c}$.
The large $\times$ marks the spontaneous zero-field magnetization at this
temperature.
For detailed discussion, see Sec.~\ref{secbind}.
The vertical, dashed line in both panels marks $m_{\rm c}$,
corresponding to the
droplet-to-slab transition in the limit $L \! \rightarrow \! \infty$, and the
vertical arrows mark $m_{\rm c}$ for the three smallest $L$ studied (from
left to right).
(a) Shows $h(m)$, highlighting the developing step discontinuity near
$m_{\rm c}$, related to the thermodynamic spinodal, and the narrow minimum
near $m_{\rm sp}$, related to the dynamic spinodal. From
bottom to top the system sizes are $L$=24, 32, 64, and 96
(the latter only for $m$$>$0.63)
(b) Shows $L\>h(m)$, highlighting the $L^{-1}$ scaling behavior of $h(m)$
in the SD region between $m_{\rm c}$ and $m_{\rm sp}$, as well
as the much slower vanishing of the minimum value, $h(m_{\rm sp})$.
The horizontal, dashed line corresponds to $L\>H_{\rm THSP}$
from Eq.~(\protect\ref{eq:hthsp2b}), and
the dotted curve is the asymptotic single-droplet result for $L\>h(m)$
obtained by applying Eq.~(\protect\ref{eqFB}) to
$F_0^{(1)}$ from Eq.~(\protect\ref{eq:f1d}).
}
\label{figFB}
\end{figure}

\begin{figure}
\caption[]{
``Spinodal phase diagram,'' showing the MD, SD, and CE
regions  in the ($1/\ln L$, $H$) plane for the two-dimensional Ising model
at 0.8$T_{\rm c}$.
The lower, solid curve is $H_{\rm THSP}$
from Eq.~(\protect\ref{eq:hthsp2b}), and the data points ($\circ$)
close to it are $H_{\rm THSP}$ from MFD1 for $L$=32 and 64.
The upper, solid curve is
$H_{\rm DSP}$ from Ref.~\protect\cite{RTMS}, obtained from a one-parameter
fit of the relation $L$$\propto$$R_0$ to MC data for $L$=64, 128, 256,
400, and 720. The short, dotted line indicates the asymptotic slope
of $H_{\rm DSP}$.
The data points represent different estimates for $H_{\rm DSP}$,
obtained from MFD1, MC in this work, $|h(m_{\rm sp})|$, and from
MC for larger systems in Ref.~\protect\cite{RTMS},
as indicated by the key in the figure.
The quantities in the upper box are all estimates for $H_{\rm DSP}$.
In all cases, error bars are only given where the statistical error is larger
than the symbol size. From right to left the data points correspond to
$L$$=$$24$, $32$, $64$, $96$, $128$, $256$, $400$, and $720$.
See detailed discussion in Sec.~\protect\ref{secbind}.
}
\label{figH}
\end{figure}


\begin{references}
\bibitem{NOTE3}
Our quantity $\Xi(T)$ is identical to
$\left[ (d \! - \! 1) / 2 m_{\rm eq}(T) \right]^{d-1} W(T)$, where $W$ is
the quantity defined in Eq.~(3) of Ref.~\protect\cite{ZA} and
also used in Ref.~\protect\cite{LZ}.
\bibitem{NOTE2}
The simulated $F_0(m)$ are very smooth, as seen in
Fig.~\protect\ref{figfm24}(a).
However, the residual noise combined with the
high resolution in $m$ ($\Delta m$=$2$$/$$L^2$) made it
necessary to calculate the numerical derivatives using a
fourth-order Savitzky-Golay smoothing filter
(subroutine {\tt savgol} from Ref.~\protect\cite{NR})
with a window on the order of the width
of the deep, narrow minimum in $h(m)$.
\end{references}
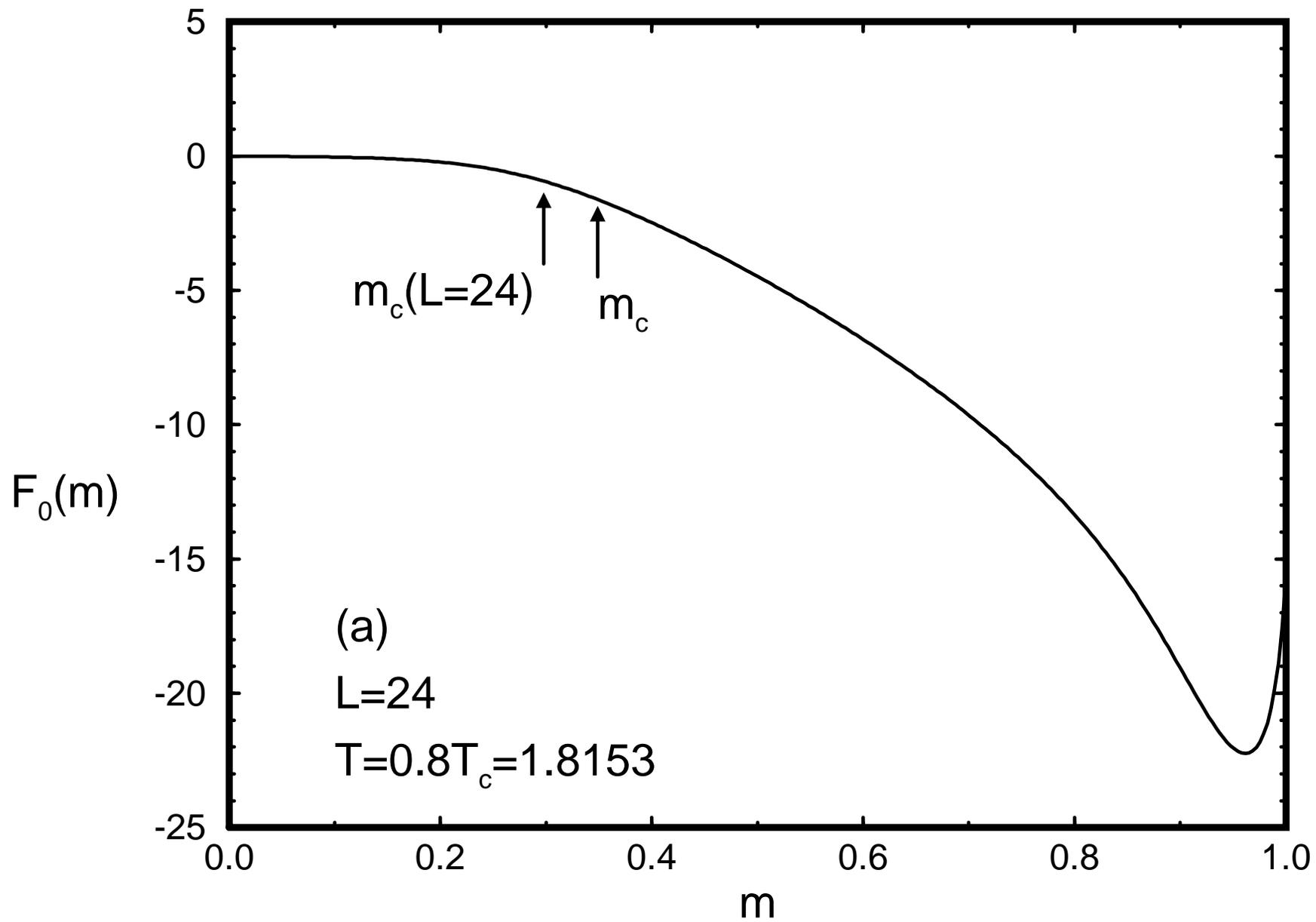
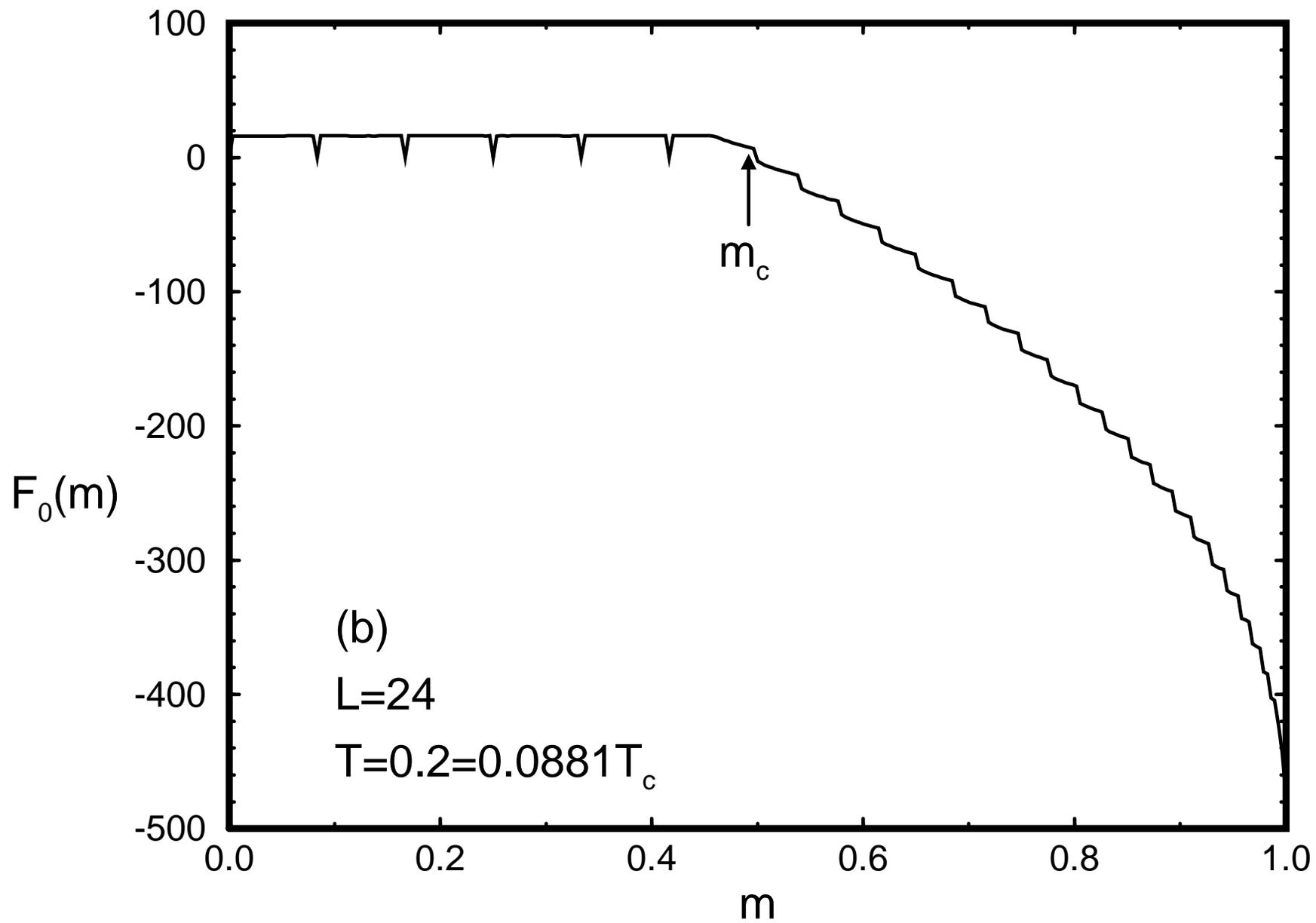
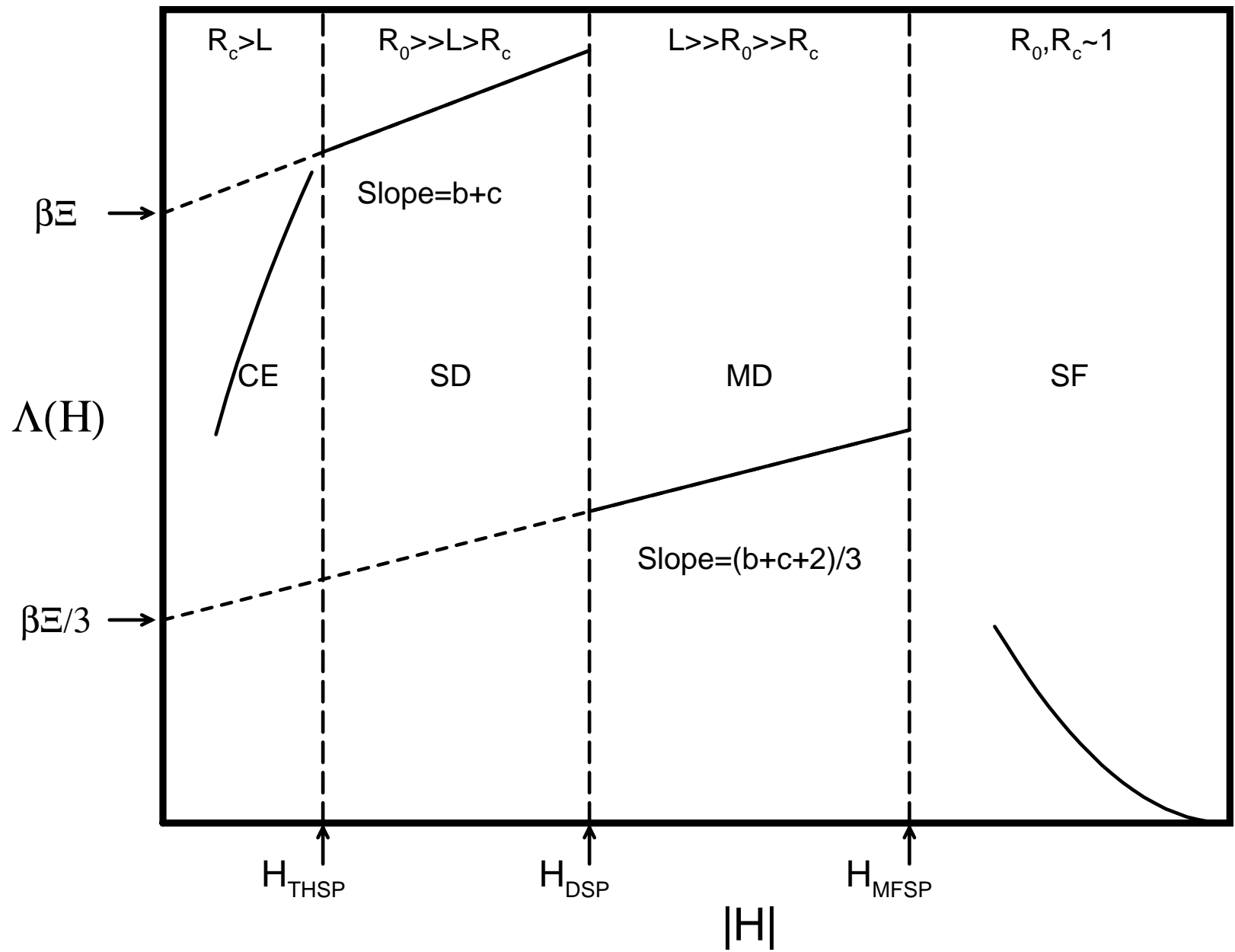
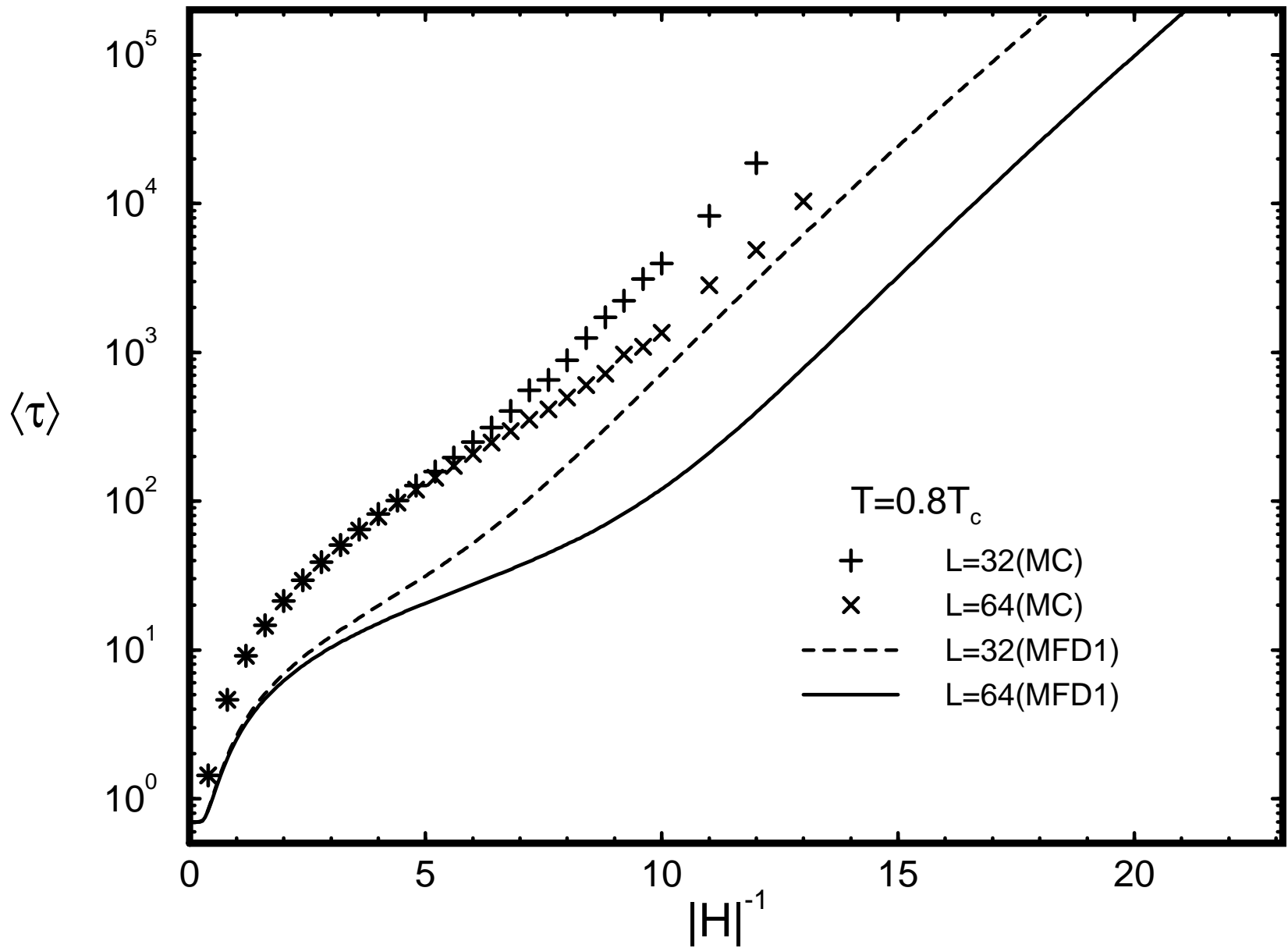
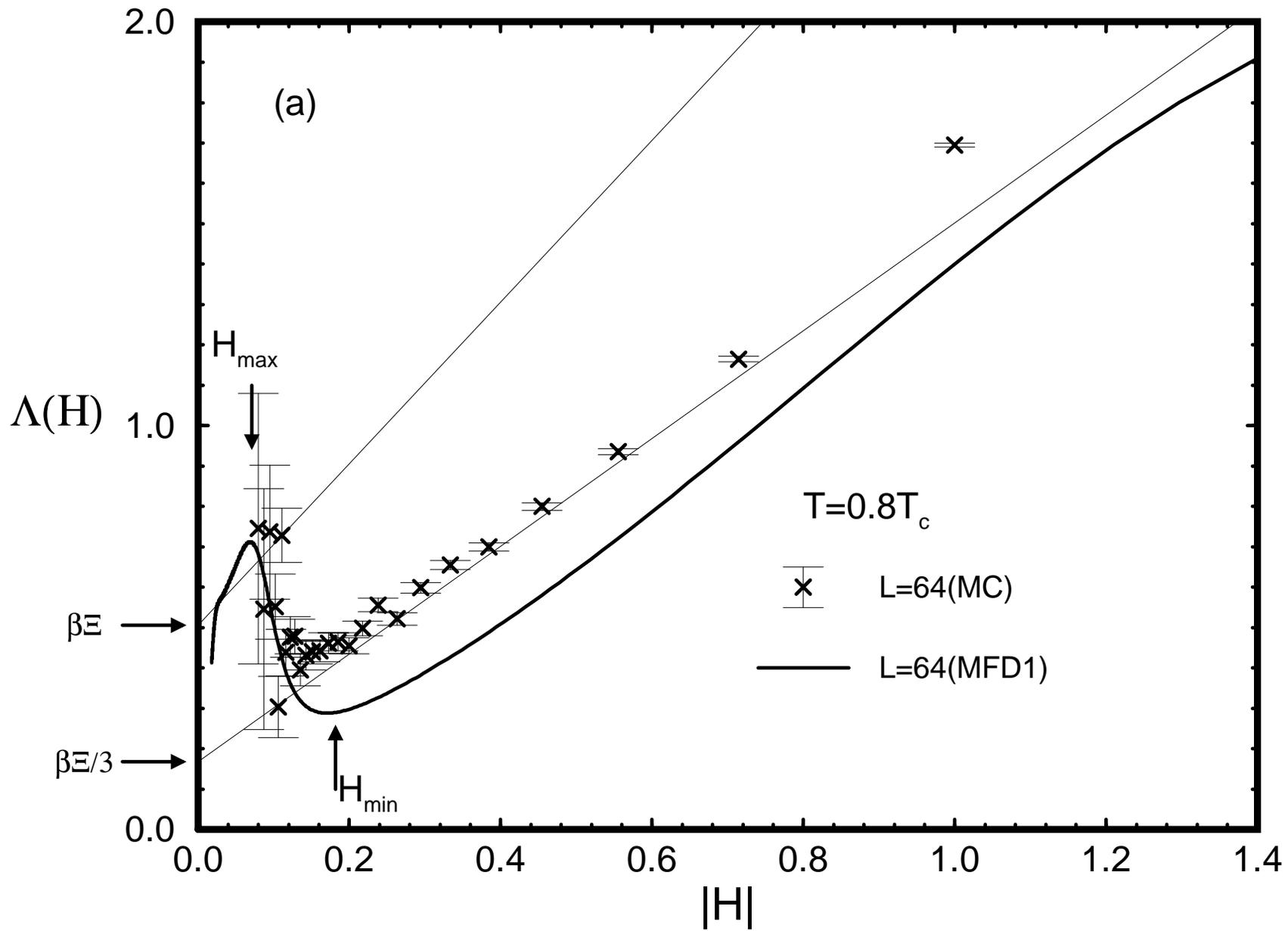
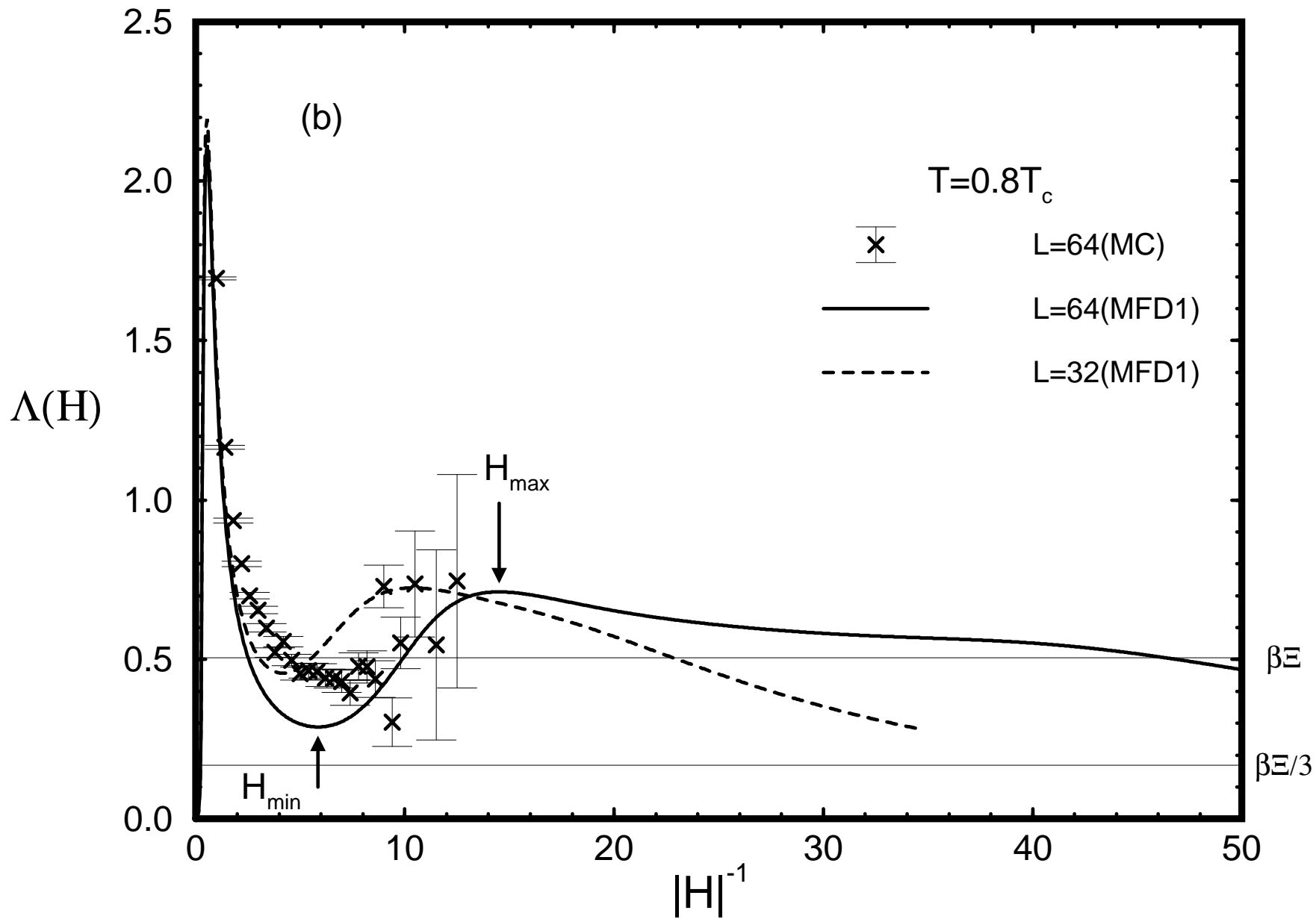
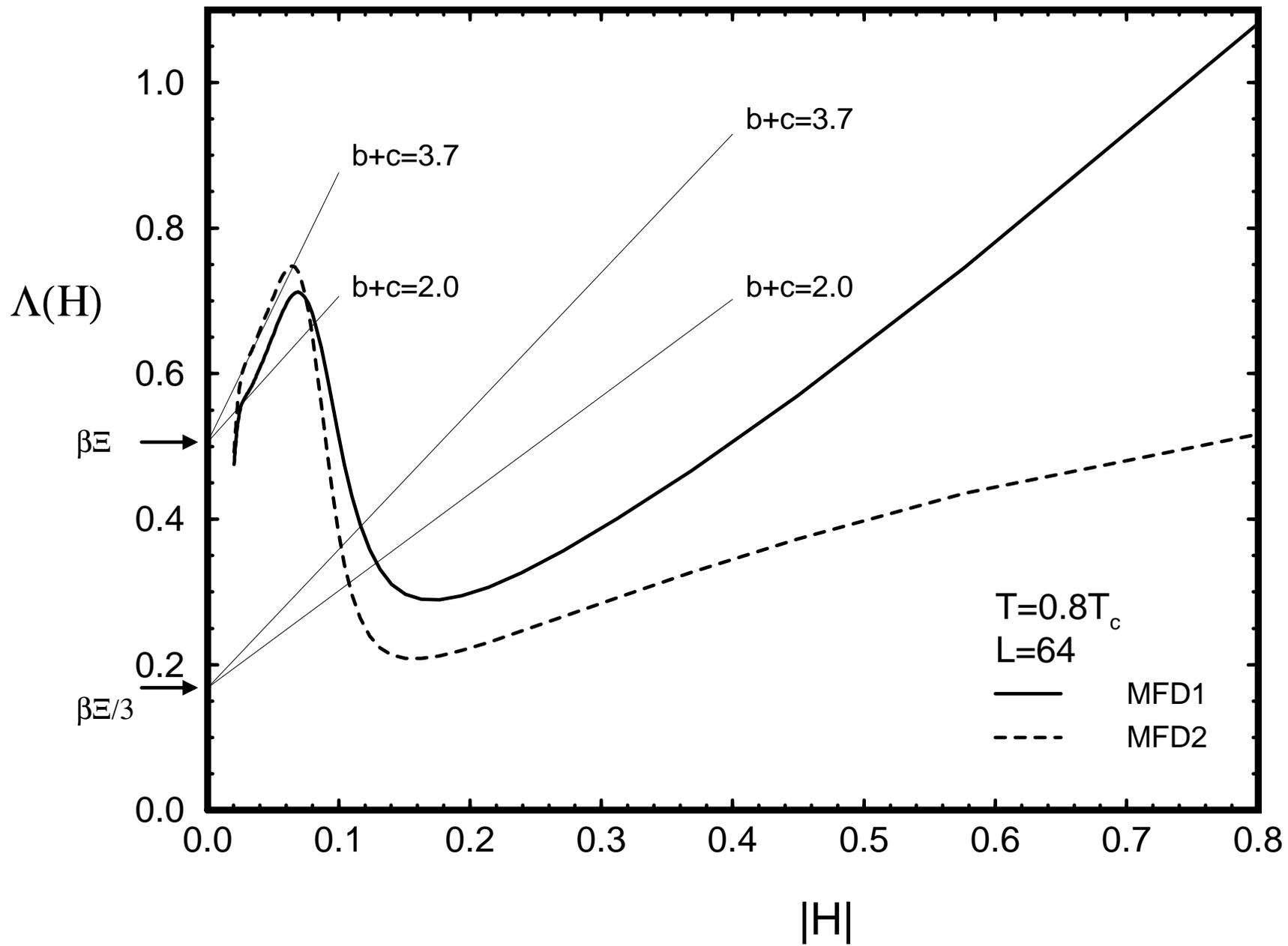
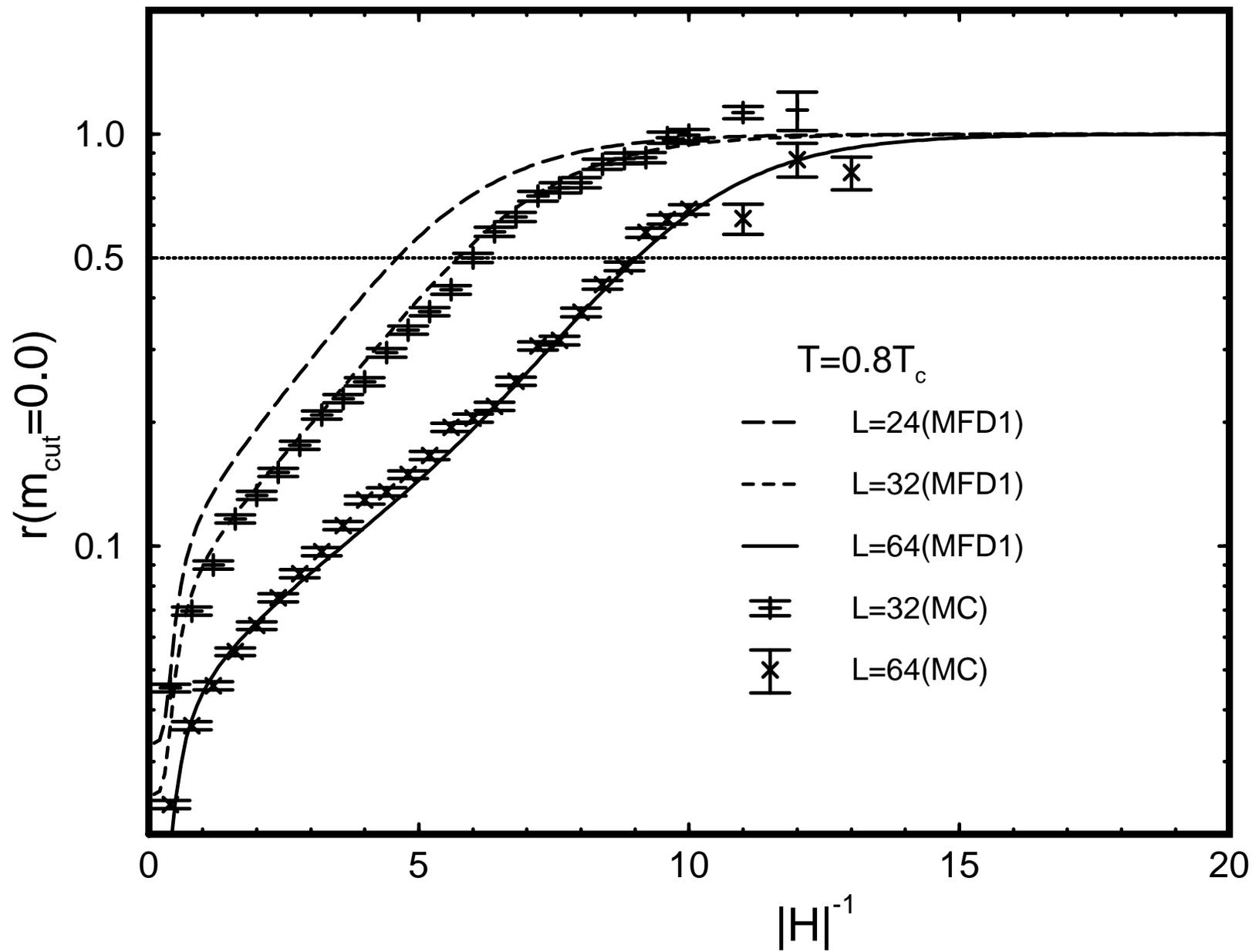
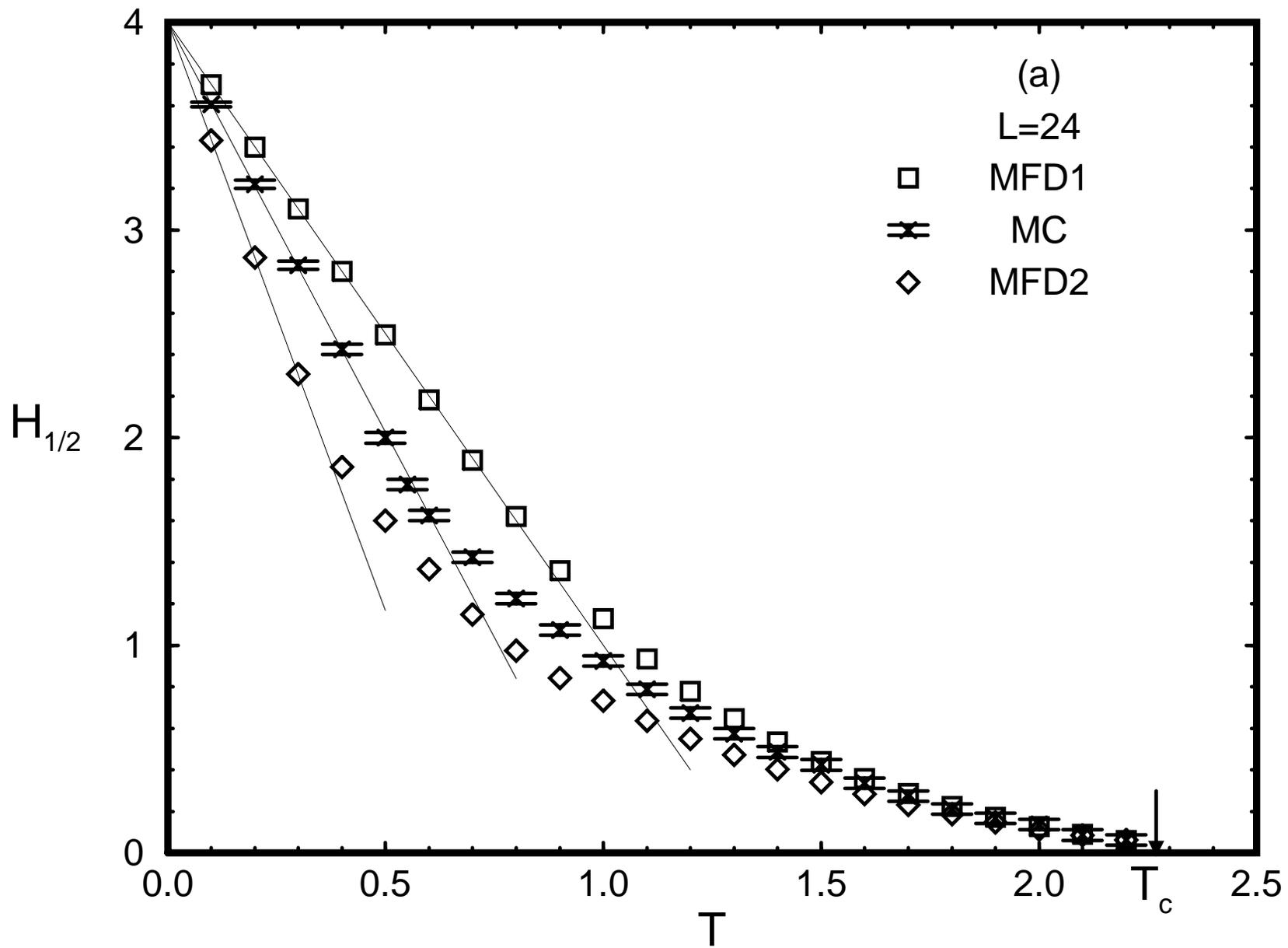
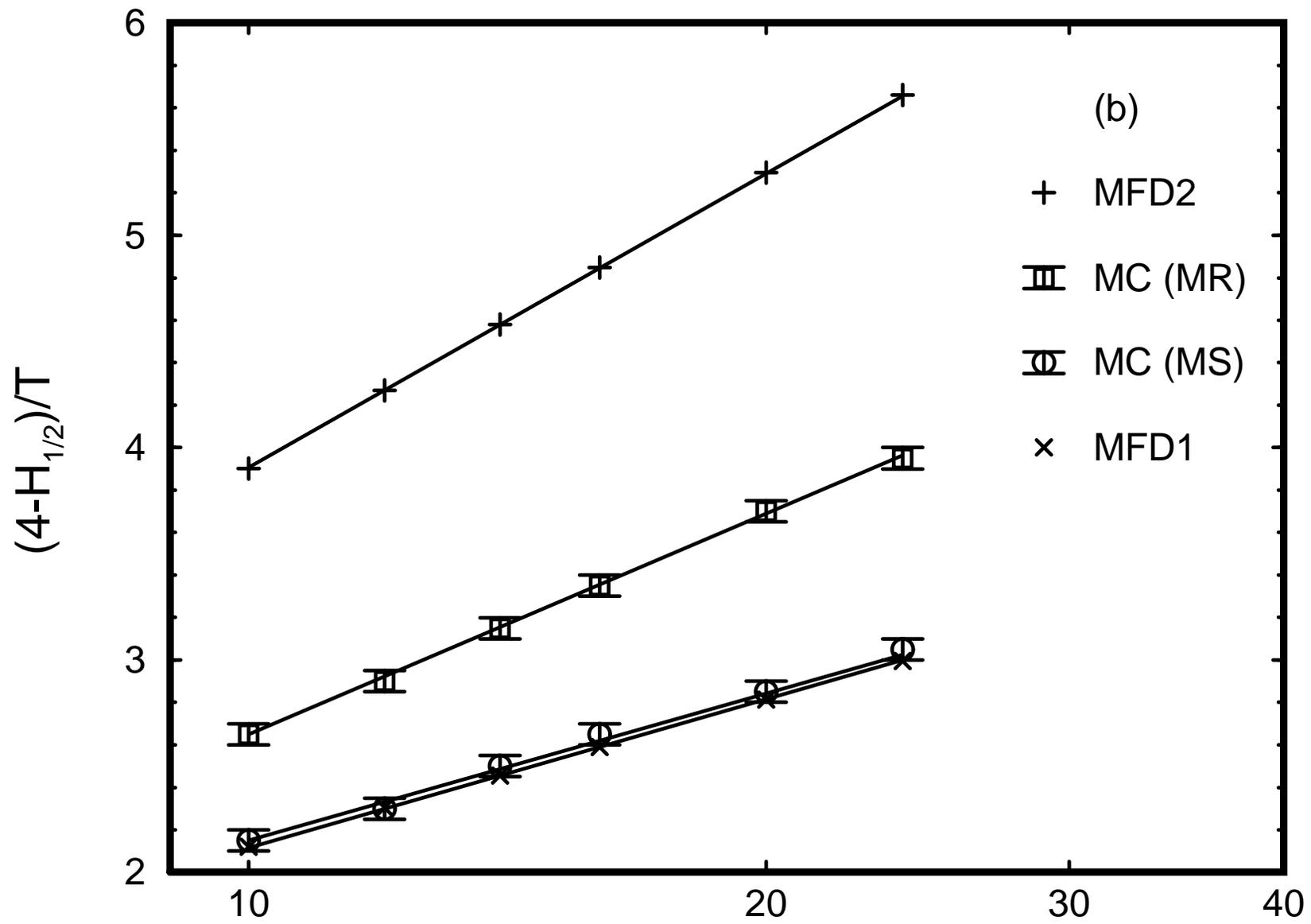
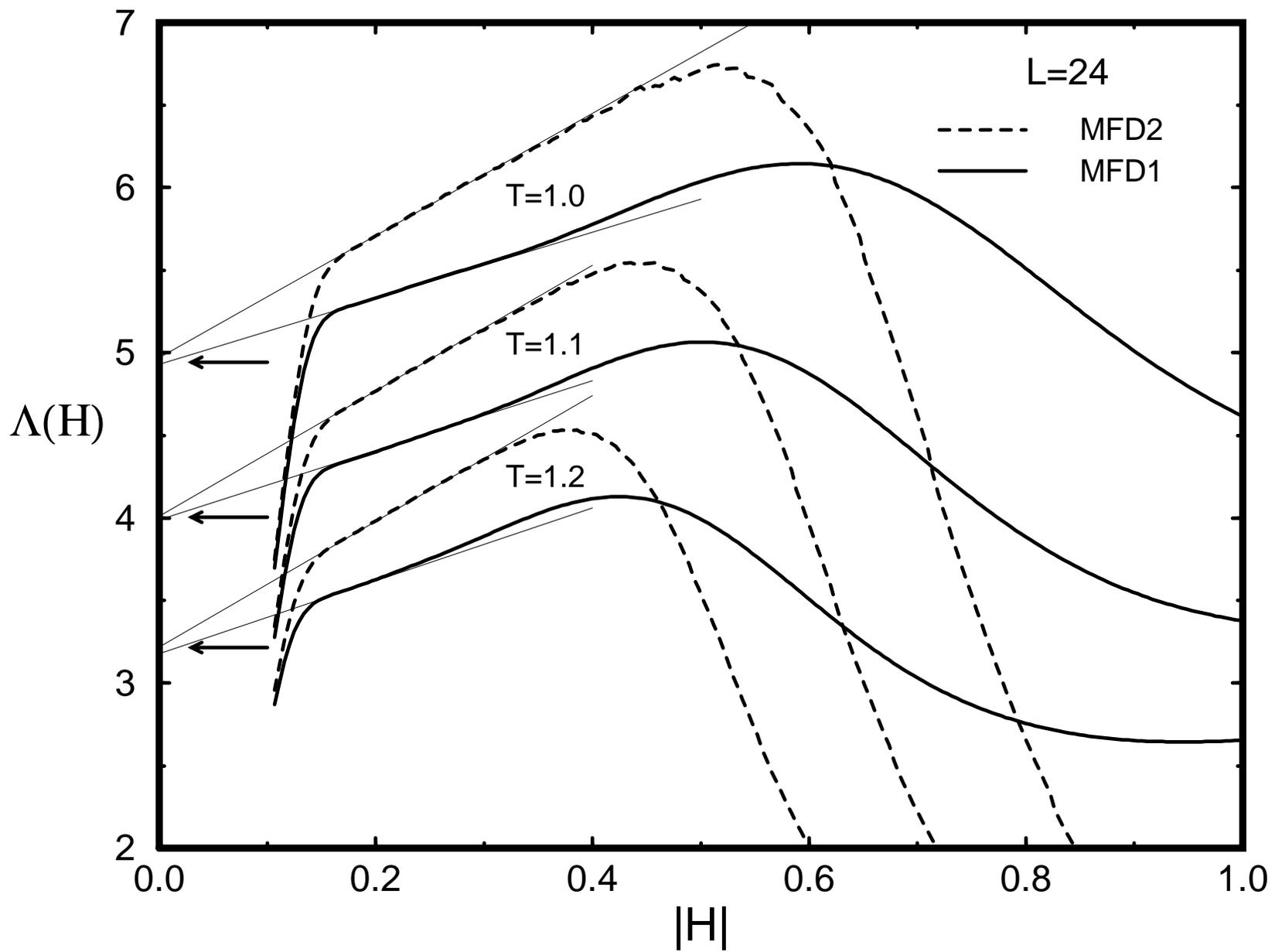
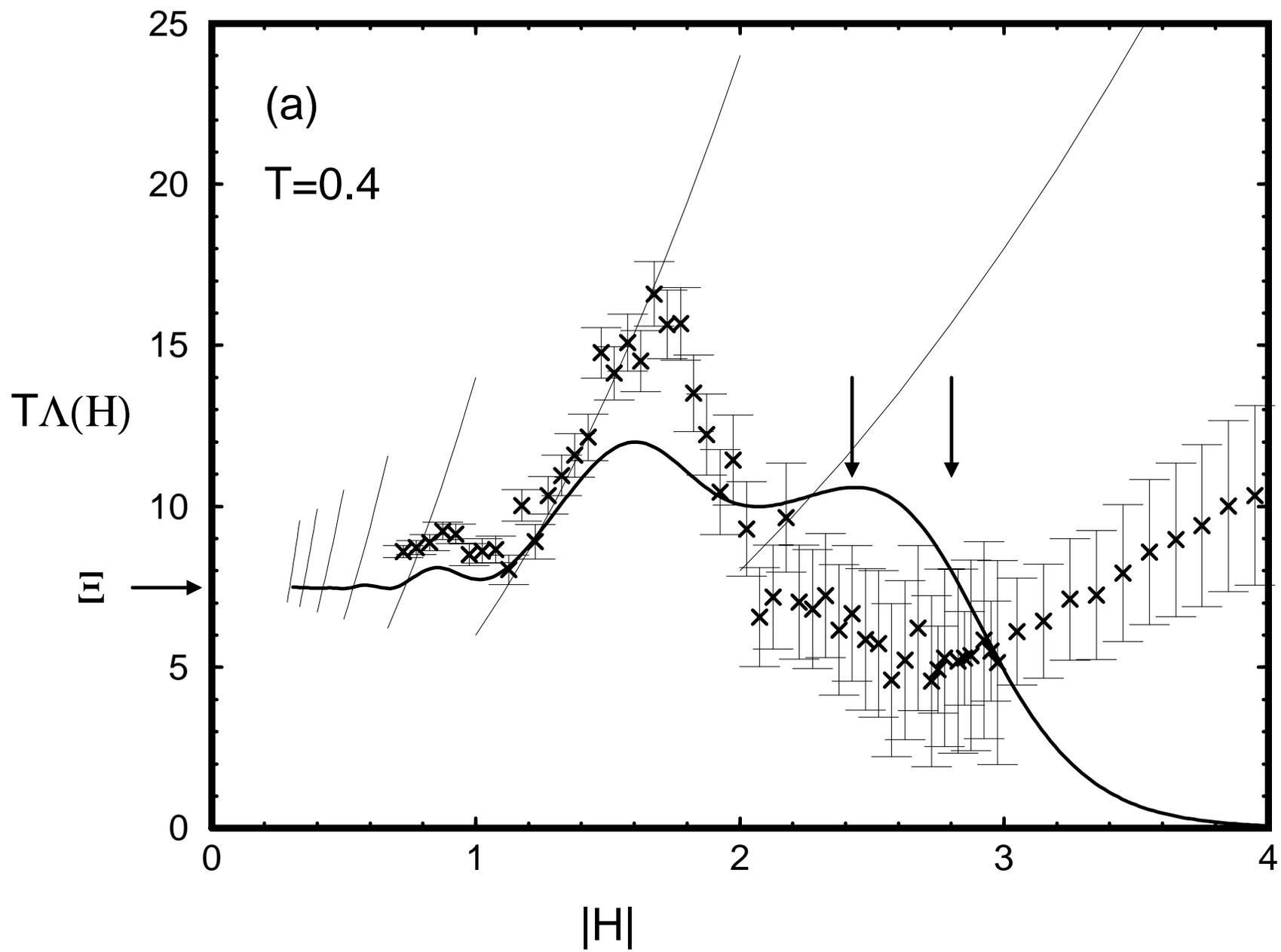
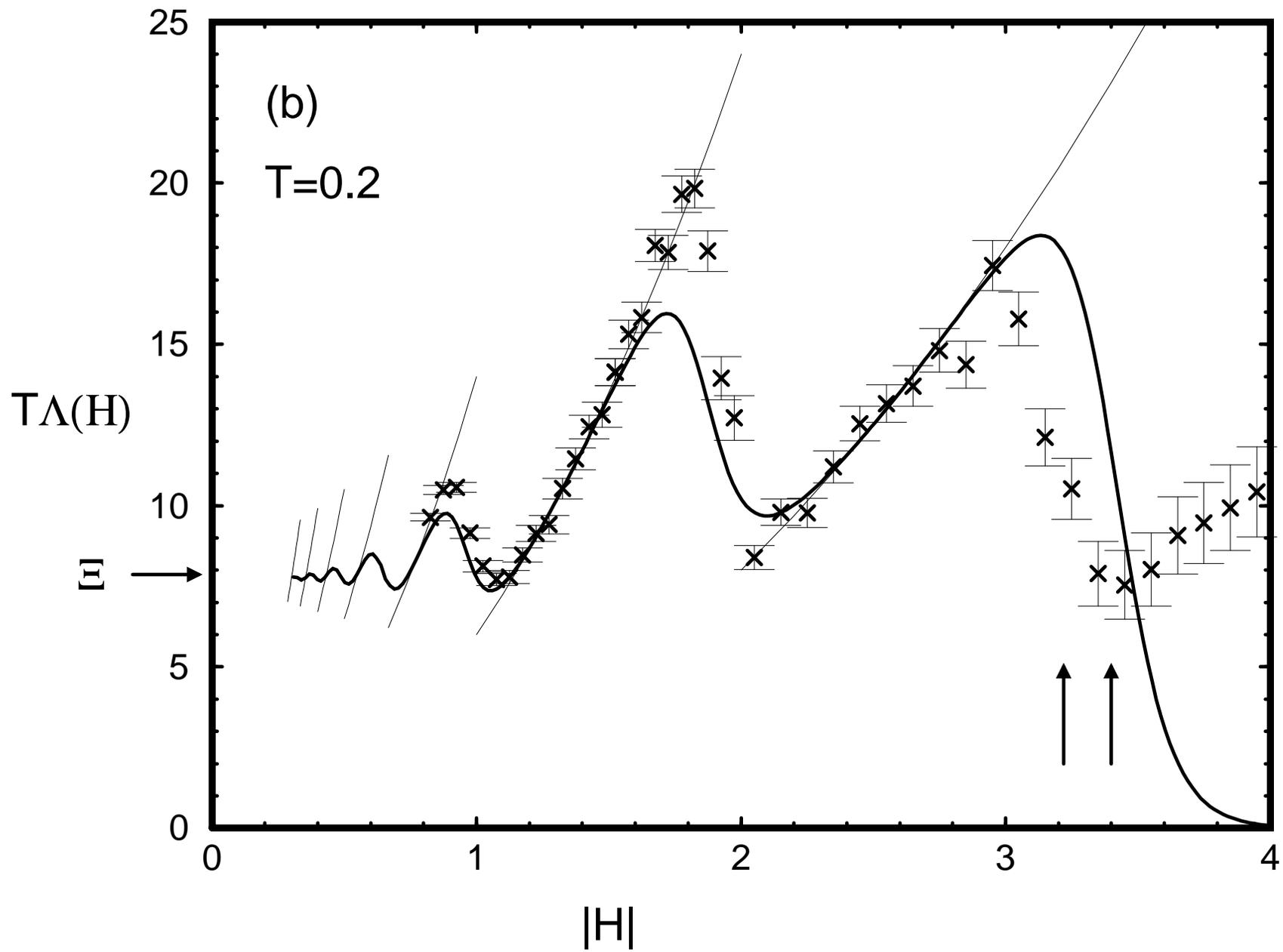
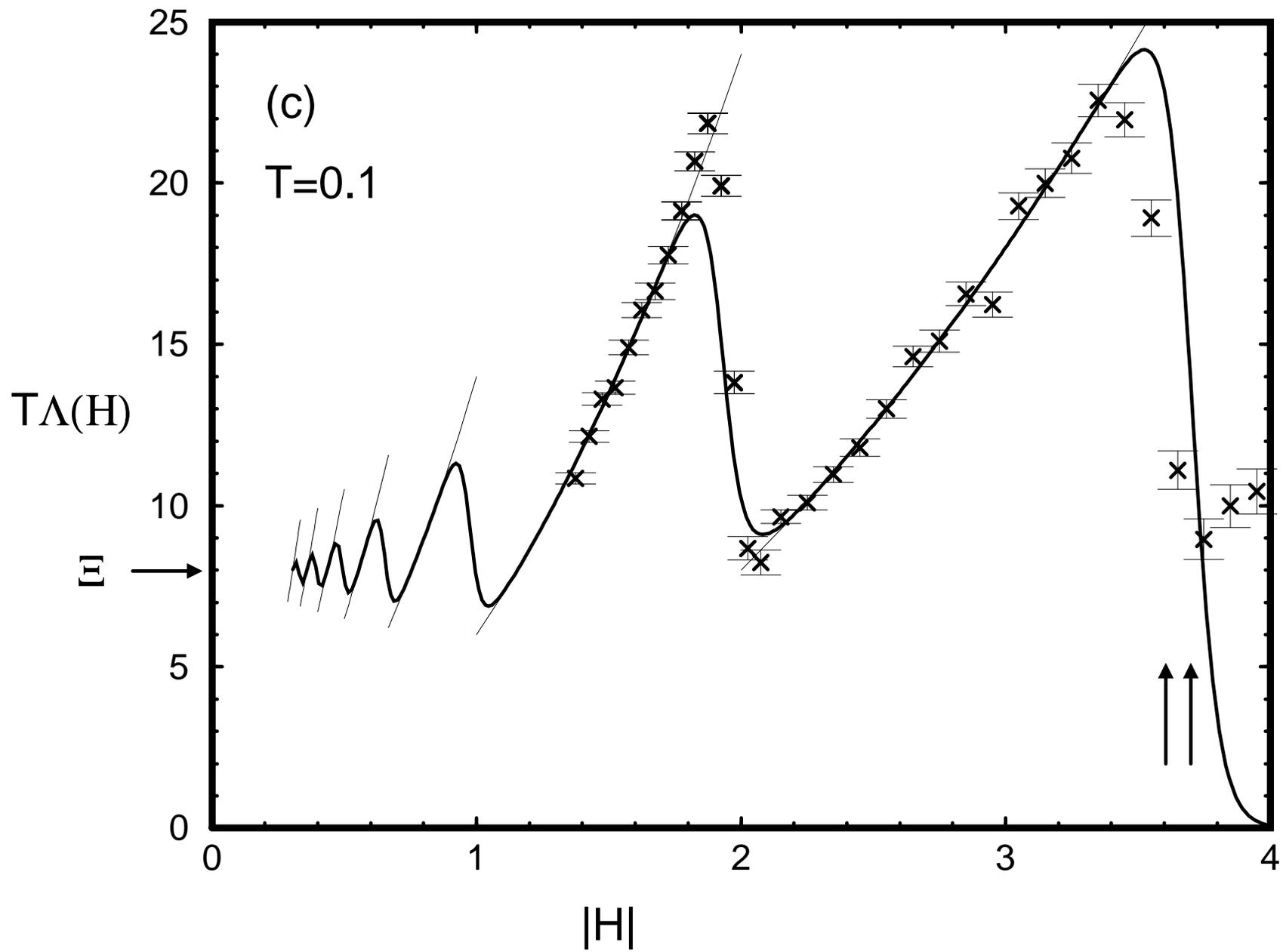
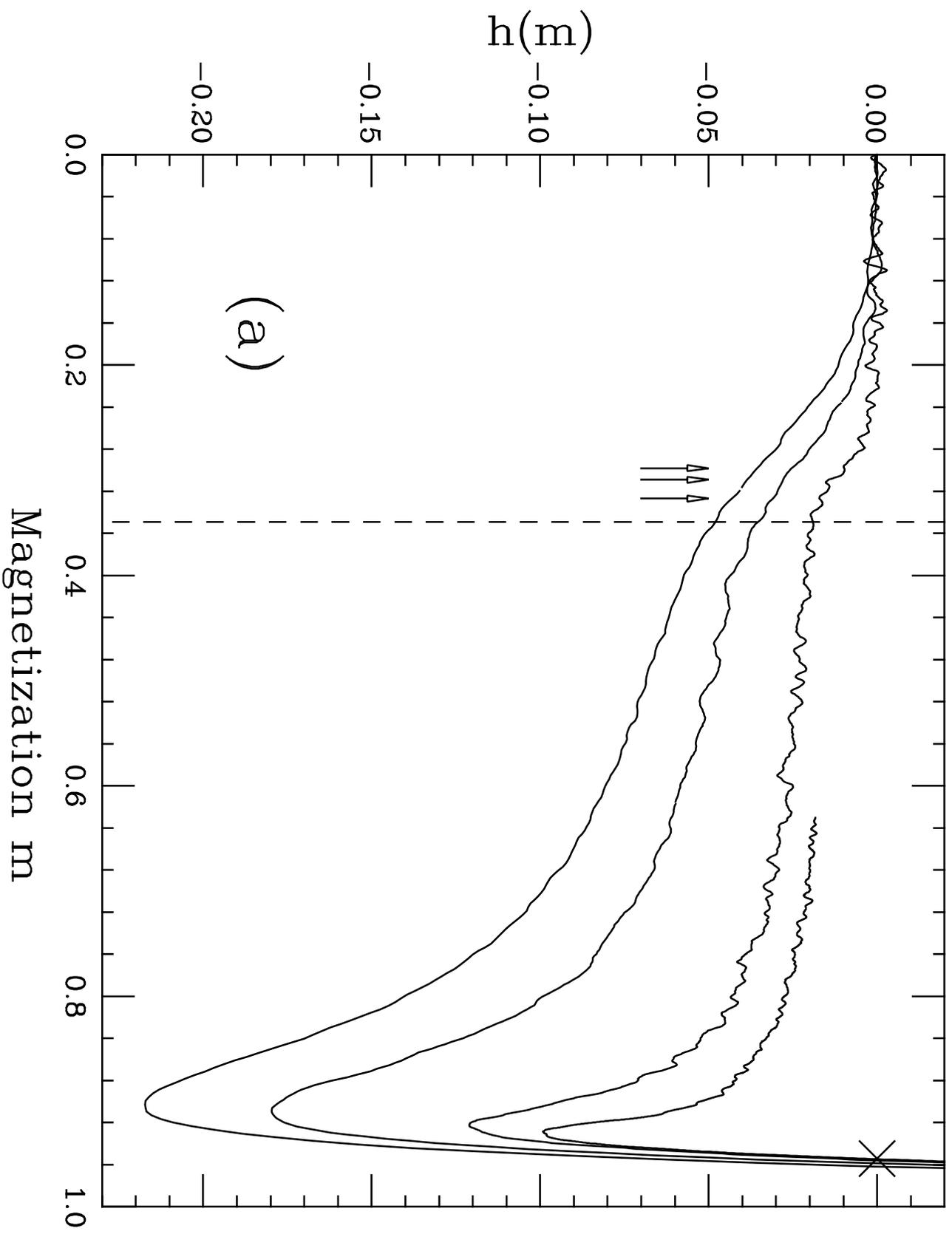
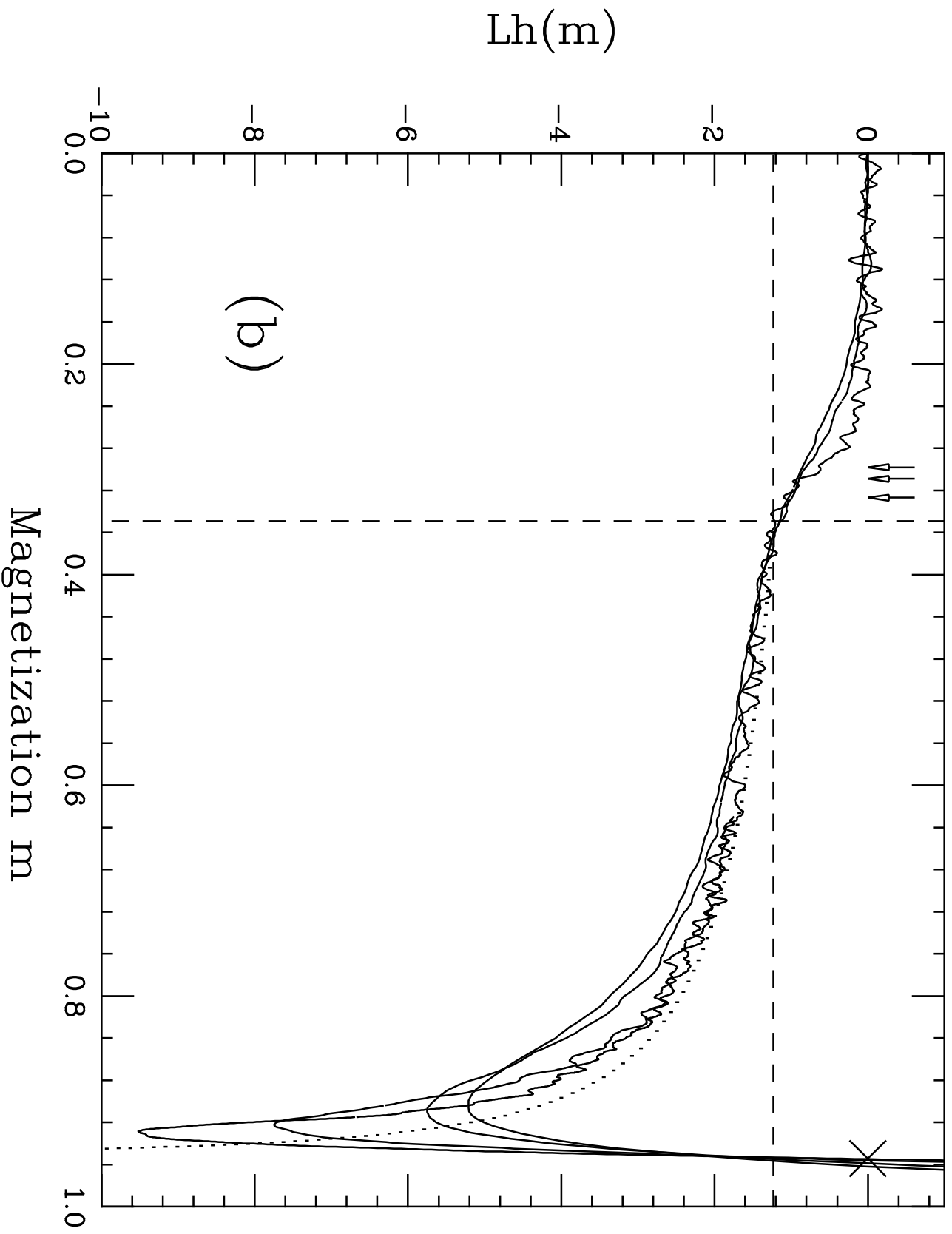
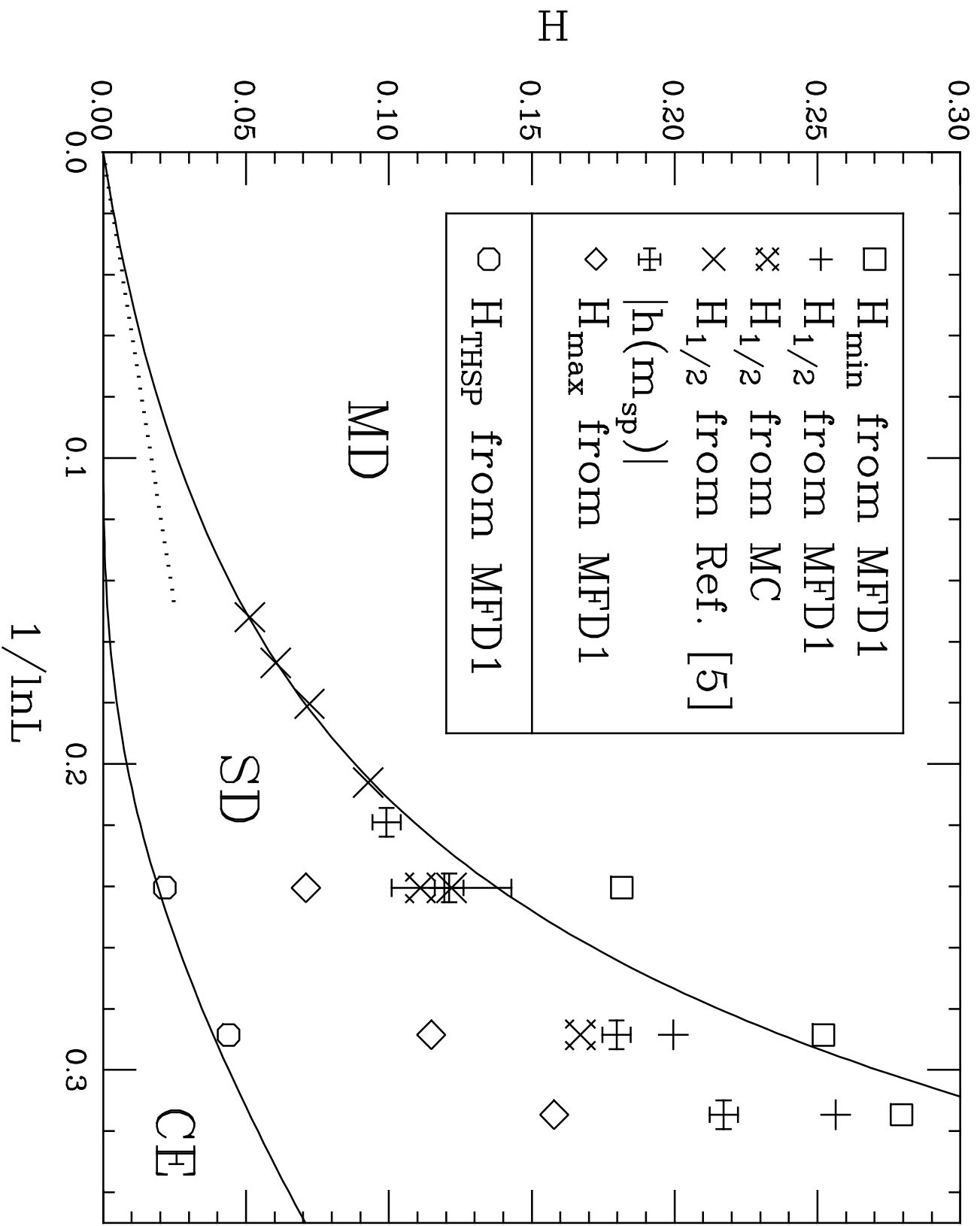
\end{document}